\documentclass[12pt]{article}
\usepackage{amsmath}
\usepackage{graphicx}
\usepackage{enumerate}
\usepackage{natbib}
\usepackage{url} 

\usepackage[section]{placeins}

\usepackage{amssymb,physics}
\usepackage{mathrsfs,bm,bbm}
\usepackage[svgnames]{xcolor}
\usepackage{tikz}
\usepackage{enumitem}
\usepackage{multirow,multicol}
\usepackage[ruled,noend]{algorithm2e}

\usepackage{amsthm}
\newtheorem{theorem}{Theorem}
\newtheorem{lemma}{Lemma}
\newtheorem{corollary}{Corollary}
\newtheorem{proposition}{Proposition}
\theoremstyle{remark}
\newtheorem{remark}{Remark}

\theoremstyle{definition}

\newtheorem{assumption}{Assumption}

\usepackage{xr-hyper}
\usepackage[bookmarksnumbered=true, colorlinks,
citecolor=ForestGreen,       
linkcolor=Blue,		         
urlcolor=Magenta,	     	 
]{hyperref}
\makeatletter
\newcommand*{\addFileDependency}[1]{
	\typeout{(#1)}
	\@addtofilelist{#1}
	\IfFileExists{#1}{}{\typeout{No file #1.}}
}
\makeatother
\newcommand*{\myexternaldocument}[1]{%
	\externaldocument{#1}%
	\addFileDependency{#1.tex}%
	\addFileDependency{#1.aux}%
}
\myexternaldocument{supp}

\graphicspath{{./art/}}

\begin{document}

\def\spacingset#1{\renewcommand{\baselinestretch}%
	{#1}\small\normalsize} \spacingset{1}

\title{\bf Functional Tensor Regression}
\author{Tongyu Li\\
	Department of Probability \& Statistics, School of Mathematical Sciences,\\ Center for Statistical Science, Peking University\\
	Fang Yao\thanks{
		Fang Yao is the corresponding author: \texttt{fyao@math.pku.edu.cn}. This research is partially supported by  the National Key R\&D Program of China (No. 2022YFA1003801), the National Natural Science Foundation of China (No. 12292981), the Newcorner Stone Science foundation through the Xplorer Prize, the LMAM and the Fundamental Research Funds for the Central Universities, Peking University (LMEQF).}\\
	Department of Probability \& Statistics, School of Mathematical Sciences,\\ Center for Statistical Science, Peking University\\
	Anru R.\ Zhang\\
	Department of Biostatistics \& Bioinformatics and\\ Department of Computer Science, Duke University}
\date{}
\maketitle

\bigskip
\begin{abstract}
Tensor regression has attracted significant attention in statistical research. This study tackles the challenge of handling covariates with smooth varying structures. We introduce a novel framework, termed functional tensor regression, which incorporates both the tensor and functional aspects of the covariate. To address the high dimensionality and functional continuity of the regression coefficient, we employ a low Tucker rank decomposition along with smooth regularization for the functional mode. We develop a functional Riemannian Gauss--Newton algorithm that demonstrates a provable quadratic convergence rate, while the estimation error bound is based on the tensor covariate dimension. Simulations and a neuroimaging analysis illustrate the finite sample performance of the proposed method.
\end{abstract}

\noindent%
\vfill

\newpage
\spacingset{1.9}

\section{Introduction}
\label{sec:intro}
Over the past few decades, tensors or multidimensional arrays have increasingly emerged in various scientific applications ranging from genomics \citep{durham2018predictd} to recommender systems \citep{bi2018multilayer} to international relations \citep{hoff2015multilinear} to computational imaging \citep{zhang2020denoising}. Such a structure is highly versatile and allows for the representation of complex data in an organized way. Moving from classic matrix-based methods to tensor-based methods, multi-linear data can be better exploited. Tensor data analysis, built on the cornerstone of tensor representation and approximation \citep{hackbusch2019tensor,kolda2009tensor}, includes low-rank tensor recovery \citep{gandy2011tensor,goldfarb2014robust}, tensor principal component analysis \citep{lu2008mpca,lu2019tensor}, tensor canonical correlation analysis \citep{kim2007tensor,luo2015tensor}, tensor classification \citep{phan2010tensor,makantasis2018tensor} and tensor regression \citep{liu2021tensor,guhaniyogi2014bayesian,guhaniyogi2017bayesian,zhou2021tensor,zhou2024broadcasted}. As far as tensor regression is concerned, a primary focus is to reveal the dependence of a scalar response on a tensor covariate, while the case of tensor response was also widely studied \citep{li2017parsimonious,sun2017store,sun2019dynamic,lock2018tensor}. 

In modern applications, tensors with a smoothly varying mode often appear within complex-structured objects. Such a mode represents features like time and spectrum and exhibits a certain level of regularity, whereas the other modes are in a tabular format. This class of tensors is called functional or dynamic \citep{bi2021tensors} and has been studied mainly from the perspective of tensor decomposition in recent years. For example, in unsupervised learning for time-varying tensors, \citet{han2023guaranteed} addressed singular value decomposition based on the theory of reproducing kernel Hilbert space, \citet{zhang2021dynamic} dealt with CP decomposition in the context of recommender systems, \citet{chen2022factor,han2024cp} investigated factor models, and \citet{sun2019dynamic} proposed a clustering method with computational efficiency. Under the supervised setup, \citet{zhou2023partially} developed a regression model with a partially observed dynamic tensor as the response, \citet{billio2024bayesian,billio2023bayesian} tackled Bayesian modeling of time series of multilayer networks via a logistic tensor-on-tensor model, and \citet{chen2024dynamic} studied the recovery of a dynamic tensor that incorporates observations of a matrix evolving smoothly over time. 

In this paper, we focus on the regression where the covariate is a functional tensor. On top of tensor decomposition, tensor regression features the difficulty of inverting a covariance tensor. Moreover, in functional tensor regression, we hope to extract the information that accumulates smoothly along the functional mode. Hence our work is closely related to functional data analysis, especially functional linear regression \citep{ramsay2005functional,ramsay1991some,hall2007methodology,crambes2009smoothing}. Taking advantage of the functional nature will also facilitate dimension reduction and data representation. 
Specific examples include:
\begin{itemize}
    \item \emph{Brain imaging}. Neuroscience is boosted by neuroimaging technologies such as functional magnetic resonance imaging, which often forms functional tensors. It is intriguing to discover the connection between neurodegenerative disorders and brain activity \citep{zhou2013tensor}.
    \item \emph{High-dimensional longitudinal data}. In longitudinal microbiome studies, a large number of bacterial taxa are measured from multiple subjects at multiple time points \citep{han2023guaranteed}. The influence of micro characteristics on the overall metrics is worth exploring.
    \item \emph{Multilayer network data}. How a network changes is often of paramount interest \citep{zhou2022network}. The adjacency matrices in multiple snapshots of dynamic networks can be stacked into a tensor \citep{jing2021community}, which provides new insights into network analysis.
\end{itemize}

To overcome the curse of dimensionality and achieve computational efficiency, most literature on tensor estimation imposes low-rank assumptions on the tensor parameter which often enhance its interpretability at the same time \citep{han2022optimal,kolda2009tensor}. However, in several cases like CP decomposition, the tensor parameter may suffer from non-identifiability and even ill-posedness \citep{zhou2021tensor}. This further complicates the functional tensor parameter whose functional mode should be distinguished. Unlike other tensor models where all modes of the coefficient tensor are treated the same way, our functional tensor regression model emphasizes the evolution along the functional mode and demands correct parameterization. Simply discretizing the functional mode may result in the loss of suitable smoothness in data, which has not been considered for tensor regression. As a consequence, new methods are needed to take into account both the tensor and functional aspects. 

Besides the formulation of functional tensor regression, it is important to solve the optimization problem corresponding to the estimation for the tensor coefficient, which is generally non-convex or NP-hard \citep{hillar2013most}. A feasible approach is to run local algorithms to refine a warm initialization \citep{chen2019non,ahmed2020tensor}, so we propose a functional Riemannian Gauss--Newton method modified from \citet{luo2023low}. Outperforming various methods, such as gradient descent, alternating minimization, and iterative hard thresholding, the Riemannian Gauss--Newton algorithm achieves a quadratic convergence rate while incurring only a moderate computation cost per iteration. We have established the estimation error bound through theoretical analysis based on a functional analog of the tensor restricted isometry property in the covariates, {which states that the norms of certain tensors are well preserved}.

We summarize the main contributions of this paper as follows. First, to our best knowledge, this is the first attempt to extend regression to the functional tensor setting. Our method utilizes the temporal modes in tensors, in contrast to the more commonly used tabular approaches. We adopt the Tucker decomposition, and circumvent the issue of non-identifiability by focusing on the overall structure. Within the proposed model, we propose a penalized least squares estimator to highlight the smoothness of changes along the functional mode. Second, when addressing the estimation challenges in functional tensor regression, particularly those involving a significant roughness penalty, we derive a new functional version of Riemannian Gauss--Newton algorithm. 
We then establish a novel quadratic convergence guarantee for the proposed algorithm. Third, our functional tensor regression framework offers a new solution to characterizing time-varying effects in tensor data analysis. We demonstrate its power through simulated and real data examples. The illustration of our method on neuroimaging reflects the effect over time and strengthens the findings in the classic literature \citet{zhou2013tensor}. 

The rest of this paper is organized as follows. Section \ref{sec:model} begins with a review of linear/tensor algebra, and then presents the model and estimation method for functional tensor regression. In Section \ref{sec:theoretical}, we establish theoretical results for the estimation error and computational complexity associated with the functional Riemannian Gauss--Newton algorithm described in Section \ref{sec:model}. The proposed method is examined by numerical performance in Section \ref{sec:numerical}. 
Technical proofs are deferred to an online Supplementary Material, while the code and data are made available in a GitHub repository (\url{https://github.com/kellty/FTReg}).

\section{Proposed Methodology}
\label{sec:model}
In this section, we introduce the functional tensor regression model together with an estimation method. To that end, we first review some notation that will be used throughout this paper. 

\subsection{Notation and Preliminaries}
Denote the Euclidean norm of any vector $\bm{v}$ by $\norm{\bm{v}}$. For any matrix $\bm{M}$, let $\sigma_{k}(\bm{M})$ be its $k$th largest singular value, and $\operatorname{SVD}(\bm{M})$ and $\operatorname{QR}(\bm{M})$ be the matrices consisting of the left singular vectors of $\bm{M}$ and the Q part of the QR decomposition of $\bm{M}$, respectively. If $\bm{M}$ is a $p$-by-$q$ matrix of rank $r$, then $\operatorname{SVD}(\bm{M})$ and $\operatorname{QR}(\bm{M})$ belong to \[ \mathbb{O}_{p,r} = \{ \bm{U} \in \mathbb{R}^{p\times r} : \bm{U}^{\top}\bm{U} = \bm{I}_{r} \} ,\] the set of $p$-by-$r$ matrices with orthonormal columns, where $\bm{U}^{\top}$ denotes the transpose of $\bm{U}$, and $\bm{I}_{r}$ denotes the $r$-by-$r$ identity matrix. For any $\bm{U} \in \mathbb{O}_{p,r}$, let $\bm{U}_{\perp} \in \mathbb{O}_{p,p-r}$ be such that $\bm{U}^{\top}\bm{U}_{\perp} = 0$, i.e., the columns of $\bm{U}$ and $\bm{U}_{\perp}$ form an orthonormal basis of $\mathbb{R}^{p}$. The Kronecker product of two matrices $\bm{M}_{1}$ and $\bm{M}_{2}$ is denoted by $\bm{M}_{1} \otimes \bm{M}_{2}$; if the $(i,j)$th entry of $\bm{M}_{1}$ is $a_{ij}$, then the $(i,j)$th block of $\bm{M}_{1} \otimes \bm{M}_{2}$ is $a_{ij}\bm{M}_{2}$.

For any tensor $\bm{\mathcal{T}} \in \mathbb{R}^{q_{0}\times q_{1}\times \dots\times q_{D}}$, let $[\bm{\mathcal{T}}]_{j_{0},j_{1},\dots,j_{D}}$ be its $(j_{0},j_{1},\dots,j_{D})$th entry. Given two tensors $\bm{\mathcal{T}}_{1},\bm{\mathcal{T}}_{2} \in \mathbb{R}^{q_{0}\times q_{1}\times \dots\times q_{D}}$ of the same order, their Frobenius inner product is defined as \[ \langle\bm{\mathcal{T}}_{1},\bm{\mathcal{T}}_{2}\rangle = \sum_{j_{0},j_{1},\dots,j_{D}} [\bm{\mathcal{T}}_{1}]_{j_{0},j_{1},\dots,j_{D}} [\bm{\mathcal{T}}_{2}]_{j_{0},j_{1},\dots,j_{D}} .\] The tensor Frobenius norm is correspondingly defined as $\norm{\bm{\mathcal{T}}}_{\mathrm{F}} = \langle\bm{\mathcal{T}},\bm{\mathcal{T}}\rangle^{1/2}$. The $d$-mode product of $\bm{\mathcal{T}} \in \mathbb{R}^{q_{0}\times q_{1}\times \dots\times q_{D}}$ with a matrix $\bm{M} \in \mathbb{R}^{p_{d}\times q_{d}}$ is denoted by $\bm{\mathcal{T}} \times_{d} \bm{M} \in \mathbb{R}^{q_{0}\times\dots\times q_{d-1}\times p_{d}\times q_{d+1}\times\dots\times q_{D}}$ whose entries are \[ [\bm{\mathcal{T}} \times_{d} \bm{M}]_{j_{0},\dots,j_{D}} = \sum_{k=1}^{q_{d}} [\bm{\mathcal{T}}]_{j_{0},\dots,j_{d-1},k,j_{d+1},\dots,j_{D}} [\bm{M}]_{j_{d},k} .\] For convenience, the tensor-matrix product along multiple modes is abbreviated as \[ \bm{\mathcal{T}} \times_{d=0}^{D} \bm{M}_{d} = \bm{\mathcal{T}} \times_{0}\bm{M}_{0} \times_{1}\bm{M}_{1} \times\dots \times_{D}\bm{M}_{D}.\] Let $\mathscr{M}_{d}$ be the operation that unfolds tensors $\bm{\mathcal{T}} \in \mathbb{R}^{q_{0}\times q_{1}\times \dots\times q_{D}}$ along mode $d$ into matrices $\mathscr{M}_{d}(\bm{\mathcal{T}}) \in \mathbb{R}^{q_{d}\times q_{-d}}$, where $q_{-d} = \prod_{e\ne d} q_{e}$, which is often termed as matricization. Formally, $[\mathscr{M}_{d}(\bm{\mathcal{T}})]_{j_{d},k} = [\bm{\mathcal{T}}]_{j_{0},j_{1},\dots,j_{D}}$ if $k = 1 + \sum_{e\ne d} (j_{e}-1) \prod_{f<e,\,f\ne d} q_{f}$. The inverse operation of $\mathscr{M}_{d}$ is denoted by $\mathscr{T}_{d} : \mathbb{R}^{q_{d}\times q_{-d}} \to \mathbb{R}^{q_{0}\times q_{1}\times \dots\times q_{D}}$, called the mode-$d$ tensorization. It can be seen that \[ \mathscr{M}_{d}(\bm{\mathcal{T}} \times_{0}\bm{M}_{0} \times\dots \times_{D}\bm{M}_{D}) = \bm{M}_{d} \mathscr{M}_{d}(\bm{\mathcal{T}}) (\bm{M}_{D}^{\top}\otimes\dots\otimes\bm{M}_{d+1}^{\top}\otimes\bm{M}_{d-1}^{\top}\otimes\dots\otimes\bm{M}_{0}^{\top}) .\] The Tucker rank of a tensor $\bm{\mathcal{T}} \in \mathbb{R}^{q_{0}\times q_{1}\times \dots\times q_{D}}$ is defined by \[ \rank_{\mathrm{Tuc}}(\bm{\mathcal{T}}) = ( \rank\mathscr{M}_{0}(\bm{\mathcal{T}}),\rank\mathscr{M}_{1}(\bm{\mathcal{T}}),\dots,\rank\mathscr{M}_{D}(\bm{\mathcal{T}}) ) .\] \citet{tucker1966some} and \citet{de2000multilinear} indicated that, if $\rank_{\mathrm{Tuc}}(\bm{\mathcal{T}}) = (r_{0},r_{1},\dots,r_{D})$, then $\bm{\mathcal{T}}$ admits the higher-order singular value decomposition: \[ \bm{\mathcal{T}} = \bm{\mathcal{S}} \times_{d=0}^{D} \bm{U}_{d} ,\] where $\bm{U}_{d} = \operatorname{SVD}\{\mathscr{M}_{d}(\bm{\mathcal{T}})\} \in \mathbb{O}_{q_{d},r_{d}}$ consists of the mode-$d$ singular vectors, and $\bm{\mathcal{S}} \in \mathbb{R}^{r_{0}\times r_{1}\times \dots\times r_{D}}$ is the core tensor. Given $\bm{r}=(r_{0},r_{1},\dots,r_{D})$, the tensor $\bm{\mathcal{T}}_{\max(\bm{r})}$ {denotes the best Tucker approximation of $\bm{\mathcal{T}}$ in the sense that $\bm{\mathcal{T}}_{\max(\bm{r})} = \bm{\mathcal{T}}\times_{d=0}^{D}(\hat{\bm{U}}_{d}\hat{\bm{U}}_{d}^{\top})$ with \[ \lVert\bm{\mathcal{T}}\times_{d=0}^{D}(\hat{\bm{U}}_{d}\hat{\bm{U}}_{d}^{\top})\rVert_{\mathrm{F}} = \max_{\bm{U}_{d}\in\mathbb{O}_{q_{d},r_{d}},\,d=0,1,\ldots,D} \norm{\bm{\mathcal{T}}\times_{d=0}^{D}(\bm{U}_{d}\bm{U}_{d}^{\top})}_{\mathrm{F}} .\]}

\subsection{Functional Tensor Regression}
We consider the regression framework where a scalar response is influenced by a functional tensor covariate, which extends tensor regression to the functional setting. To model the functional mode, we adopt the concepts of the functional linear regression model introduced by \citet{ramsay1991some}. Its classical version relates a functional covariate $x$ to the response $y$ via $E(y\mid x) = \int_{\mathbb{T}} x(t) \beta(t) \dd{t}$, where $\beta$ is the unknown coefficient function and $\mathbb{T}$ is the continuous domain on which $x$ and $\beta$ are defined. This motivates our functional tensor model: 
\begin{equation*}
y = \int_{\mathbb{T}} \langle \bm{\mathcal{X}}(t) , \bm{\mathcal{B}}(t) \rangle \dd{t} + \varepsilon ,
\end{equation*}
where $y$ is a continuous response, $\bm{\mathcal{X}}(\bm{\cdot})$ is a functional tensor covariate, $\bm{\mathcal{B}}(\bm{\cdot})$ is an unknown functional tensor coefficient, and $\varepsilon$ is a zero-mean error. The parameter $\bm{\mathcal{B}}(\bm{\cdot})$ is functional and thus can be evaluated at any time point, which exhibits a smooth time-varying effect. The functions $\bm{\mathcal{X}}(\bm{\cdot})$ and $\bm{\mathcal{B}}(\bm{\cdot})$ are defined on the interval $\mathbb{T}$, take values in $\mathbb{R}^{p_{1}\times \dots\times p_{D}}$, and are required to satisfy some regularity condition. 
The functional tensor $\bm{\mathcal{X}}(\bm{\cdot})$ admits the Karhunen--Lo\`eve decomposition 
\begin{equation}\label{eq:KL_decomp}
\bm{\mathcal{X}}(\bm{\cdot}) = \sum_{k=1}^{\infty} \bm{\varXi}_{k} \varphi_{k}(\bm{\cdot}) , 
\end{equation}
where $\varphi_{k}$'s form an orthonormal basis of the space $L^{2}(\mathbb{T})$ of square-integrable functions, and $\bm{\varXi}_{k} = \int_{\mathbb{T}} \bm{\mathcal{X}}(t) \varphi_{k}(t) \dd{t}$ are uncorrelated random tensors. The concrete conditions on $\bm{\mathcal{X}}(\bm{\cdot})$ are listed in Section \ref{sec:theoretical}. 
In addition, the entries of $\bm{\mathcal{B}}(\bm{\cdot})$ lie in some Sobolev space $W^{m,2}(\mathbb{T})$, that is, $\bm{\mathcal{B}}(\bm{\cdot})$ is $m$-times differentiable and the entries of its $m$th derivative $\bm{\mathcal{B}}^{(m)}(\bm{\cdot})$ are square-integrable. 
We assume that the training data $(y_{i},\bm{\mathcal{X}}_{i})$, $i=1,\ldots,n$, consists of $n$ independent copies of $(y,\bm{\mathcal{X}})$, and that $\bm{\mathcal{X}}_{i}$'s are measured 
only over a discrete grid $t_{1}<\dots<t_{p_0}$ with observational noise, i.e., 
\begin{equation}\label{eq:obs}
\bm{\mathcal{X}}_{ij} = \bm{\mathcal{X}}_{i}(t_{j}) + \bm{\mathcal{E}}_{ij} , 
\end{equation}
where $\bm{\mathcal{E}}_{ij}$ is a $p_{1}\times \dots\times p_{D}$ tensor with entries of zero mean and finite variance. 
To facilitate global retrieval, we assume further that there exists a constant $C_{0} > 0$ for which $C_{0}^{-1} \le p_{0} (t_{j+1}-t_{j}) \le C_{0}$, $j=0,1,\ldots,p_{0}$, where $t_{0}$ and $t_{p_{0}+1}$ are the endpoints of $\mathbb{T}$. {The aligned measurement points $t_j$'s allow us to focus on a tensor structure of the unknown coefficient $\bm{\mathcal{B}}(\bm{\cdot})$, which further leads to an affordable computational burden. It is nontrivial to extend this framework to irregularly spaced  observations, and we leave it as future work.}
In what follows, we write $\bm{y}$ and $\bm{\varepsilon}$ for the vectors of $y_{i}$'s and $\varepsilon_{i}$'s, respectively, where $\varepsilon_{i} = y_{i} - \int_{\mathbb{T}} \langle \bm{\mathcal{X}}_{i}(t) , \bm{\mathcal{B}}(t) \rangle \dd{t}$. 

Based on the above model, our goal is to estimate $\bm{\mathcal{B}}(\bm{\cdot})$. As is shown in \citet{crambes2009smoothing}, the influence of $\bm{\mathcal{X}}(\bm{\cdot})$ at each measurement point $t_{j}$ could be quantified by $\bm{\mathcal{B}}_{j} = \bm{\mathcal{B}}(t_{j})$, which leads to the idea of interpolating $\bm{\mathcal{B}}_{j}$'s to recover $\bm{\mathcal{B}}(\bm{\cdot})$. Hence, for simplicity, $\bm{\mathcal{B}}(\bm{\cdot})$ is supposed to be an interpolant through $(t_{j},\bm{\mathcal{B}}_{j})$, $j=1,\ldots,p_{0}$, which turns out to be unique if natural splines are used. See \citet[Chapter 5]{eubank1999nonparametric} for smoothing splines. Let $\bm{\psi}(\bm{\cdot}) = (\psi_{1}(\bm{\cdot}),\dots,\psi_{p_0}(\bm{\cdot}))^{\top}$ be a basis of the space of natural splines with order $2m$ and knots $t_{1},\dots,t_{p_0}$. With $\bm{\varTheta} \in \mathbb{R}^{p_{0}\times p_{1}\times \dots\times p_{D}}$ and $\bm{\Psi} \in \mathbb{R}^{p_{0}\times p_{0}}$ being such that $[\bm{\varTheta}]_{j_{0},j_{1},\dots,j_{D}} = [\bm{\mathcal{B}}_{j_{0}}]_{j_{1},\dots,j_{D}}$ and $[\bm{\Psi}]_{j,k} = \psi_{k}(t_{j})$, the coefficient of interest is 
\begin{equation}\label{eq:interp}
\bm{\mathcal{B}}(\bm{\cdot}) = \bm{\varTheta} \times_{0} \{ \bm{\psi}(\bm{\cdot})^{\top} (\bm{\Psi}^{\top}\bm{\Psi})^{-1} \bm{\Psi}^{\top} \} ,
\end{equation}
{where the $0$-mode product corresponds to spline interpolation in the functional mode.}
{This formulation allows for easy manipulation of unknown parameters, and we can mitigate the curse of dimensionality by imposing additional structure on the tensor $\bm{\varTheta}$; otherwise, the unprocessed vectorization of $\bm{\varTheta}$ would lead to prohibitive computation in an ultra-high dimensional space.}

Then, it suffices to solve the estimation problem for $\bm{\varTheta}$ whose zeroth mode is essentially functional. There is a significant difference in functional tensor regression compared with the tabular tensor regression: one must consider time intervals and smoothness to achieve estimation efficiency. For a given value of tuning parameter $\rho > 0$, we obtain an estimator $\widehat{\bm{\varTheta}}$ by minimizing a penalized squared loss $L(\bm{\varTheta}) + \rho J(\bm{\varTheta})$ in some suitable space of $\bm{\varTheta}$. Here 
\begin{equation}\label{eq:mse}
\begin{aligned}
L(\bm{\varTheta}) &= (2n)^{-1}\sum_{i=1}^{n} \Big( y_{i} - \sum_{j=1}^{p_0} \langle \bm{\mathcal{X}}_{ij} , \bm{\mathcal{B}}_{j} \rangle \Delta t_{j} \Big)^{2} \\
&= (2n)^{-1}\sum_{i=1}^{n} \big( y_{i} - \langle \bm{\mathcal{Z}}_{i} , \bm{\varTheta} \rangle \big)^{2} 
= (2n)^{-1} \norm{ \bm{y} - \mathscr{Z}\bm{\varTheta} }^{2} 
\end{aligned}
\end{equation}
for some interval sizes $\Delta t_{1}, \dots, \Delta t_{p_0} > 0$, where $\bm{\mathcal{Z}}_{i}$ is the $p_{0}\times p_{1}\times \dots\times p_{D}$ tensor such that 
\begin{equation*}
[\bm{\mathcal{Z}}_{i}]_{j_{0},j_{1},\dots,j_{D}} = \Delta t_{j_0} [\bm{\mathcal{X}}_{ij_{0}}]_{j_{1},\dots,j_{D}} ,
\end{equation*}
and $\mathscr{Z}$ is the linear map that sends $\bm{\varTheta}$ to the vector $( \langle \bm{\mathcal{Z}}_{1} , \bm{\varTheta} \rangle , \dots , \langle \bm{\mathcal{Z}}_{n} , \bm{\varTheta} \rangle )^{\top}$. We shall choose $\Delta t_{j} = (t_{j+1}-t_{j-1})/2$, which {is the mean of $t_{j+1}-t_{j}$ and $t_{j}-t_{j-1}$, and} satisfies the quasi-uniform property that 
\begin{equation}\label{eq:spacing}
C_{0}^{-1} \le p_{0} \Delta t_{j} \le C_{0} .
\end{equation}

The penalty term $J(\bm{\varTheta})$ is designed to control the roughness of $\bm{\mathcal{B}}(\bm{\cdot})$ and ensure the existence of a unique penalized least-squares solution, which requires careful treatment in tensor regression \citep{zhou2014regularized}. 
Motivated by the method developed in \citet{crambes2009smoothing}, we take 
\begin{equation}\label{eq:penalty}
J(\bm{\varTheta}) = \int_{\mathbb{T}} \norm{\bm{\mathcal{B}}^{(m)}(t)}_{\mathrm{F}}^{2} \dd{t} + \sum_{j=1}^{p_0} \Delta t_{j} \bigg\lVert \sum_{k=1}^{m}\tilde{\bm{\mathcal{A}}_{k}}t_{j}^{k-1} \bigg\rVert_{\mathrm{F}}^{2} 
\end{equation}
with $(\tilde{\bm{\mathcal{A}}_{1}},\dots,\tilde{\bm{\mathcal{A}}_{m}})$ being the minimizer of $\sum_{j=1}^{p_0} \Delta t_{j} \lVert \bm{\mathcal{B}}_{j}-\sum_{k=1}^{m}\bm{\mathcal{A}}_{k}t_{j}^{k-1} \rVert_{\mathrm{F}}^{2}$ over all $m$-tuples of $p_{1}\times \dots\times p_{D}$ tensors $(\bm{\mathcal{A}}_{1},\dots,\bm{\mathcal{A}}_{m})$. The first term {penalizing roughness} appears quite often in regularization approach and reads $\langle \bm{\varTheta} \times_{0} \bm{\Omega}_{m}, \bm{\varTheta} \rangle$ by \eqref{eq:interp}, where 
\begin{equation}\label{eq:penalty-der}
\bm{\Omega}_{m} = \bm{\Psi} (\bm{\Psi}^{\top}\bm{\Psi})^{-1} \Big\{ \int_{\mathbb{T}} \bm{\psi}^{(m)}(t)\bm{\psi}^{(m)}(t)^{\top} \dd{t} \Big\} (\bm{\Psi}^{\top}\bm{\Psi})^{-1} \bm{\Psi}^{\top},
\end{equation}
a positive semi-definite $p_{0}$-by-$p_{0}$ matrix. 
{The second term complements the first term so that $J(\bm{\varTheta})$ is a positive definite quadratic form of $\bm{\varTheta}$.}
Let $\bm{G}$ be the $p_{0}$-by-$m$ matrix with $[\bm{G}]_{j,k} = t_{j}^{k-1}$ and $\bm{\Delta}$ be the diagonal matrix constructed from $\Delta t_{j}$'s. Denote \[ \bm{P}_{m} = \bm{G} (\bm{G}^{\top}\bm{\Delta}\bm{G})^{-1}\bm{G}^{\top}\bm{\Delta} ,\] a (not necessarily orthogonal) projection matrix. Then the second term on the right-hand side of \eqref{eq:penalty} can be written as
\begin{equation*}
\langle (\bm{\varTheta} \times_{0} \bm{P}_{m}) \times_{0} \bm{\Delta} , \bm{\varTheta} \times_{0} \bm{P}_{m} \rangle = \langle \bm{\varTheta} \times_{0} (\bm{\Delta}\bm{P}_{m}) , \bm{\varTheta} \rangle . 
\end{equation*}
Combining this and \eqref{eq:penalty-der}, the penalty \eqref{eq:penalty} becomes
\begin{equation*}
J(\bm{\varTheta}) = \langle \bm{\varTheta} \times_{0} \bm{A}_{m} , \bm{\varTheta} \rangle = \langle \mathscr{A} \bm{\varTheta}, \bm{\varTheta} \rangle. 
\end{equation*}
Here \[ \bm{A}_{m} = \bm{\Omega}_{m}+\bm{P}_{m}^{\top}\bm{\Delta}\bm{P}_{m} = \bm{\Omega}_{m}+\bm{\Delta}\bm{P}_{m} \] and $\mathscr{A}$ is the linear map sending $\bm{\varTheta}$ to $\bm{\varTheta} \times_{0} \bm{A}_{m}$. To see the positive definiteness of $\bm{A}_{m}$, notice that the null space of $\bm{\Omega}_{m}$ is {the column space of $\bm{G}$, and if $\bm{u}=\bm{G}\bm{v}\in\mathbb{R}^{p_0}$ for some $\bm{v}\in\mathbb{R}^{m}$,} \[ \bm{u}^{\top}\bm{A}_{m}\bm{u} = \bm{u}^{\top}\bm{\Delta}\bm{u} \ge (C_{0}p_{0})^{-1} \norm{\bm{u}}^{2} \] by \eqref{eq:spacing}. On the other hand, the behavior of the eigenvalues of $\bm{\Omega}_{m}$ has been well studied \citep[see, e.g.,][]{utreras1983natural}, where the smallest nonzero one $\sigma_{p_{0}-m}(\bm{\Omega}_{m})$ is of order $p_{0}^{-1}$. Consequently, there exists some constant $c_{m} > 0$ such that the smallest eigenvalue of $\bm{A}_{m}$ has the lower bound 
\begin{equation}\label{eq:pen_low_bound}
\sigma_{p_0}(\bm{A}_{m}) \ge c_{m}p_{0}^{-1} .
\end{equation}

We end this subsection with a note that the parameter is no longer infinite-dimensional, but some characteristics of nonparametric methods will still appear. The spline interpolation relies on the observation points, whose number $p_{0}$ is not bounded, so we need to trade off between goodness of fit and model parsimony with respect to the functional mode. Meanwhile, the dimensionality of tabular modes {plays a prominent role when modeling the whole tensor}. Therefore, we are confronted with an exceptionally flexible model and need careful analysis.

\subsection{Functional Riemannian Gauss--Newton Algorithm}
\label{subsec:RGN}
To fit the functional tensor model, the fidelity to data and the smoothness of the functional parameter are both addressed. Now the loss function can be written in a ridge form:
\begin{equation}\label{eq:loss}
L(\bm{\varTheta}) + \rho J(\bm{\varTheta}) = (2n)^{-1} \norm{ \bm{y} - \mathscr{Z}\bm{\varTheta} }^{2} + \rho \langle \mathscr{A} \bm{\varTheta} , \bm{\varTheta} \rangle ,
\end{equation}
incorporating \eqref{eq:mse} and \eqref{eq:penalty}. The minimization problem associated with \eqref{eq:loss} is fairly complicated due to its extremely high dimensionality, so it is imperative to reduce the parameter size to a manipulable level. To this end, we assume that the true parameter belongs to \[ \mathbb{M}_{\bm{r}} = \{ \bm{\varTheta} \in \mathbb{R}^{p_{0}\times p_{1}\times \dots\times p_{D}} : \rank_{\mathrm{Tuc}}(\bm{\varTheta}) = \bm{r} \} \] for some $\bm{r} = (r_{0},r_{1},\dots,r_{D})$. 
Note that Tucker decomposition is flexible for allowing different numbers of factors along each mode.
The flexibility in selecting different ranks for various tensor modes is advantageous when the tensor data are dimensionally skewed, a scenario that commonly appears in neuroimaging data \citep{li2018tucker}. 
The set $\mathbb{M}_{\bm{r}}$ is a smooth manifold of dimension $\prod_{d=0}^{D}r_{d}+\sum_{d=0}^{D}r_{d}(p_{d}-r_{d})$ embedded into $\mathbb{R}^{p_{0}\times p_{1}\times \dots\times p_{D}}$ \citep{koch2010dynamical,uschmajew2013geometry}. 
{Note that the intrinsic dimension also has important implications for the theoretical analysis, especially with respect to the requirement on the sample size $n$.}
Riemannian optimization techniques shed light on the problem of minimizing \eqref{eq:loss} on $\mathbb{M}_{\bm{r}}$; see \citet{absil2009optimization,boumal2023introduction} for an introduction. In a Riemannian optimization procedure, an iteration step is typically carried out by updating the point on the tangent space and then retracting it to the manifold. Next, we derive a functional Riemannian Gauss--Newton scheme. {Compared to \citet{luo2023low}, the novelty of our work on functional tensor regression lies in the appropriate treatment of the penalty term, which enables fast computation as established in Proposition~\ref{prop:LAL} below.}

We start by describing the tangent space of $\mathbb{M}_{\bm{r}}$. Let the Tucker decomposition of $\bm{\varTheta}\in\mathbb{M}_{\bm{r}}$ be $\bm{\varTheta} = \bm{\mathcal{S}} \times_{d=0}^{D} \bm{U}_{d}$ where $\bm{\mathcal{S}} \in \mathbb{R}^{r_{0}\times r_{1}\times \dots\times r_{D}}$ and $\bm{U}_{d} \in \mathbb{O}_{p_{d},r_{d}}$, and let \[ \bm{V}_{d} = \operatorname{QR}\{\mathscr{M}_{d}(\bm{\mathcal{S}})^{\top}\} \in \mathbb{O}_{r_{-d},r_{d}} \] where $r_{-d} = \prod_{e\ne d} r_{e}$. By \citet[Theorem 2.1]{koch2010dynamical}, the elements of the tangent space $\mathrm{T}_{\bm{\varTheta}}\mathbb{M}_{\bm{r}}$ of $\mathbb{M}_{\bm{r}}$ at $\bm{\varTheta}$ are 
\begin{equation*}
 \bm{\mathcal{C}} \times_{d=0}^{D} \bm{U}_{d} + \sum_{d=0}^{D} \mathscr{T}_{d}(\bm{D}_{d}\bm{V}_{d}^{\top}) \times_{d} \bm{U}_{d\perp} \times_{e\ne d} \bm{U}_{e} 
= \bm{\mathcal{C}} \times_{d=0}^{D} \bm{U}_{d} + \sum_{d=0}^{D} \mathscr{T}_{d}(\bm{U}_{d\perp}\bm{D}_{d}\bm{W}_{d}^{\top}) , \\
\end{equation*}
$\bm{\mathcal{C}} \in \mathbb{R}^{r_{0}\times r_{1}\times \dots\times r_{D}}$, $\bm{D}_{d} \in \mathbb{R}^{(p_{d}-r_{d})\times r_{d}} ,\ d=0,1,\ldots,D$. Here $\bm{W}_{d}$ is defined to be
\begin{equation}\label{eq:row_space_matrix}
\bm{W}_{d} = (\bm{U}_{D}\otimes\dots\otimes\bm{U}_{d+1}\otimes\bm{U}_{d-1}\otimes\dots\otimes\bm{U}_{0}) \bm{V}_{d} , 
\end{equation}
which corresponds to the row space of $\mathscr{M}_{d}(\bm{\varTheta})$. Thus, the tangent space $\mathrm{T}_{\bm{\varTheta}}\mathbb{M}_{\bm{r}}$ can be parameterized by \[ \mathbb{D} = \mathbb{R}^{r_{0}\times r_{1}\times \dots\times r_{D}} \times \mathbb{R}^{(p_{0}-r_{0})\times r_{0}} \times \mathbb{R}^{(p_{1}-r_{1})\times r_{1}} \times \dots \times \mathbb{R}^{(p_{D}-r_{D})\times r_{D}} .\] In conjunction with \citet[Lemma 3.1]{koch2010dynamical}, \citet{luo2023low} showed that the projection operator $\mathscr{P}_{\bm{\varTheta}}$ from $\mathbb{R}^{p_{0}\times p_{1}\times \dots\times p_{D}}$ onto $\mathrm{T}_{\bm{\varTheta}}\mathbb{M}_{\bm{r}}$ can be decomposed into $\mathscr{P}_{\bm{\varTheta}} = \mathscr{R}_{\bm{\varTheta}} \mathscr{R}_{\bm{\varTheta}}^{*}$, where
\begin{equation}\label{eq:proj_TM}
\begin{aligned}
\mathscr{P}_{\bm{\varTheta}} : & \mathbb{R}^{p_{0}\times p_{1}\times \dots\times p_{D}} \to \mathrm{T}_{\bm{\varTheta}}\mathbb{M}_{\bm{r}} \subset \mathbb{R}^{p_{0}\times p_{1}\times \dots\times p_{D}} \\
& \bm{\varUpsilon} \mapsto \bm{\varUpsilon}\times_{d=0}^{D}(\bm{U}_{d}\bm{U}_{d}^{\top}) + \sum_{d=0}^{D} \mathscr{T}_{d}\{\bm{U}_{d\perp}\bm{U}_{d\perp}^{\top}\mathscr{M}_{d}(\bm{\varUpsilon})\bm{W}_{d}\bm{W}_{d}^{\top}\} , 
\end{aligned}
\end{equation}
\begin{equation}\label{eq:extension_TM}
\mathscr{R}_{\bm{\varTheta}} : \big(\bm{\mathcal{C}},(\bm{D}_{d})_{d=0}^{D}\big) \in \mathbb{D} \mapsto \bm{\mathcal{C}} \times_{d=0}^{D} \bm{U}_{d} + \sum_{d=0}^{D} \mathscr{T}_{d}(\bm{U}_{d\perp}\bm{D}_{d}\bm{W}_{d}^{\top}) \in \mathrm{T}_{\bm{\varTheta}}\mathbb{M}_{\bm{r}} , 
\end{equation}
\begin{equation}\label{eq:contraction_TM}
\mathscr{R}_{\bm{\varTheta}}^{*} : \bm{\varUpsilon}\in\mathbb{R}^{p_{0}\times p_{1}\times \dots\times p_{D}} \mapsto \big( \bm{\varUpsilon}\times_{d=0}^{D}\bm{U}_{d}^{\top} , \{\bm{U}_{d\perp}^{\top}\mathscr{M}_{d}(\bm{\varUpsilon})\bm{W}_{d}\}_{d=0}^{D} \big) \in\mathbb{D} . 
\end{equation}
The linear operators $\mathscr{R}_{\bm{\varTheta}}$ and $\mathscr{R}_{\bm{\varTheta}}^{*}$ represent extension and contraction, respectively. 
{The image of $\mathscr{R}_{\bm{\varTheta}}$ contains generic elements of $\mathrm{T}_{\bm{\varTheta}}\mathbb{M}_{\bm{r}}$, and $\mathscr{R}_{\bm{\varTheta}}^{*}$ is constructed such that $\mathscr{R}_{\bm{\varTheta}} \mathscr{R}_{\bm{\varTheta}}^{*} = \mathscr{P}_{\bm{\varTheta}}$.}

Given the current iterate $\bm{\varTheta}^{k}$, the objective function \eqref{eq:loss} evaluated at $\mathscr{R}_{\bm{\varTheta}^{k}}\big(\bm{\mathcal{C}},(\bm{D}_{d})_{d=0}^{D}\big)$ in place of $\bm{\varTheta}$ is 
\begin{equation*}
(2n)^{-1} \norm{ \bm{y} - \mathscr{Z}\mathscr{R}_{\bm{\varTheta}^{k}}\big(\bm{\mathcal{C}},(\bm{D}_{d})_{d=0}^{D}\big) }^{2} + \rho \langle \mathscr{A} \mathscr{R}_{\bm{\varTheta}^{k}}\big(\bm{\mathcal{C}},(\bm{D}_{d})_{d=0}^{D}\big) , \mathscr{R}_{\bm{\varTheta}^{k}}\big(\bm{\mathcal{C}},(\bm{D}_{d})_{d=0}^{D}\big) \rangle ,
\end{equation*}
where the number of parameters is equal to $\dim\mathbb{M}_{\bm{r}} = \prod_{d=0}^{D}r_{d}+\sum_{d=0}^{D}r_{d}(p_{d}-r_{d})$ and becomes computationally feasible. The solution minimizing it is explicitly 
\begin{equation}\label{eq:update_half}
\check{\bm{\varTheta}}^{k+1} = \mathscr{R}_{\bm{\varTheta}^{k}} ( \mathscr{R}_{\bm{\varTheta}^{k}}^{*}\mathscr{Z}^{*}\mathscr{Z}\mathscr{R}_{\bm{\varTheta}^{k}} + n \rho \mathscr{R}_{\bm{\varTheta}^{k}}^{*}\mathscr{A}\mathscr{R}_{\bm{\varTheta}^{k}} )^{-1} \mathscr{R}_{\bm{\varTheta}^{k}}^{*}\mathscr{Z}^{*}\bm{y} , 
\end{equation}
where $\mathscr{Z}^{*}$ is the adjoint operator of $\mathscr{Z}$, i.e., $\bm{y} \mapsto \sum_{i=1}^{n} y_{i} \bm{\mathcal{Z}}_{i}$. 
The regularization ensures that there is no problem with matrix invertibility when $\rho>0$. 
For expressing $\mathscr{R}_{\bm{\varTheta}^{k}}^{*}\mathscr{A}\mathscr{R}_{\bm{\varTheta}^{k}}$, we have the following proposition.
\begin{proposition}\label{prop:LAL}
Let $\mathscr{R}_{\bm{\varTheta}}$ and $\mathscr{R}_{\bm{\varTheta}}^{*}$ be defined in \eqref{eq:extension_TM} and \eqref{eq:contraction_TM}, and $\mathscr{A}$ be the linear map $\bm{\varTheta} \mapsto \bm{\varTheta} \times_{0} \bm{A}$. Then for $\bm{\mathcal{C}} \in \mathbb{R}^{r_{0}\times r_{1}\times \dots\times r_{D}}$, $\bm{D}_{d} \in \mathbb{R}^{(p_{d}-r_{d})\times r_{d}} ,\ d=0,1,\ldots,D$, 
\begin{align*}
\mathscr{R}_{\bm{\varTheta}}^{*}\mathscr{A}\mathscr{R}_{\bm{\varTheta}}\big(\bm{\mathcal{C}},(\bm{D}_{d})_{d=0}^{D}\big) = \big( 
& \bm{\mathcal{C}}\times_{0}(\bm{U}_{0}^{\top}\bm{A}\bm{U}_{0})+\mathscr{T}_{0}(\bm{D}_{0}\bm{V}_{0}^{\top})\times_{0}(\bm{U}_{0}^{\top}\bm{A}\bm{U}_{0\perp}) , \\
& \bm{U}_{0\perp}^{\top}\bm{A}\bm{U}_{0}\mathscr{M}_{0}(\bm{\mathcal{C}})\bm{V}_{0}+\bm{U}_{0\perp}^{\top}\bm{A}\bm{U}_{0\perp}\bm{D}_{0} , \\
& \big[ \mathscr{M}_{d}\{\mathscr{T}_{d}(\bm{D}_{d}\bm{V}_{d}^{\top})\times_{0}(\bm{U}_{0}^{\top}\bm{A}\bm{U}_{0})\}\bm{V}_{d} \big]_{d=1}^{D} \big) . 
\end{align*}
\end{proposition}

Then we map $\check{\bm{\varTheta}}^{k+1} \in \mathrm{T}_{\bm{\varTheta}^{k}}\mathbb{M}_{\bm{r}}$ in \eqref{eq:update_half} back to the manifold $\mathbb{M}_{\bm{r}}$. An ideal choice is the truncated higher-order singular value decomposition (T-HOSVD) which constructs lower Tucker rank approximations; see \citet{vannieuwenhoven2012new}. The detailed procedure of T-HOSVD is given in Appendix~\ref{sec:T-HOSVD}. Denote the T-HOSVD operation by $\mathcal{H}_{\bm{r}} : \mathbb{R}^{p_{0}\times p_{1}\times \dots\times p_{D}} \to \mathbb{M}_{\bm{r}}$. It is known that $\mathcal{H}_{\bm{r}}$ satisfies the quasi-projection property \citep[Theorem 10.2]{hackbusch2019tensor}
\begin{equation}\label{eq:T-HOSVD=proj}
\norm{\bm{\varUpsilon}-\mathcal{H}_{\bm{r}}(\bm{\varUpsilon})}_{\mathrm{F}} \le (D+1)^{1/2} \norm{\bm{\varUpsilon}-\bm{\varUpsilon}'}_{\mathrm{F}} ,\quad \bm{\varUpsilon}\in\mathbb{R}^{p_{0}\times p_{1}\times \dots\times p_{D}} ,\ \bm{\varUpsilon}'\in\mathbb{M}_{\bm{r}} .
\end{equation}
In view of this, we update the solution to \eqref{eq:loss} by 
\begin{equation}\label{eq:update}
\bm{\varTheta}^{k+1} = \mathcal{H}_{\bm{r}}(\check{\bm{\varTheta}}^{k+1}) .
\end{equation}

The above functional Riemannian Gauss--Newton steps are summarized in Algorithm \ref{alg:est} below. {Note that the sample size $n$ and parameter dimension $p_0,p_1,\dots,p_D$ are given by data, while the rank $\bm{r}$ and the tuning parameter $\rho$ needs selection, for which we suggest a generalized cross-validation detailed in Section~\ref{sec:numerical}.}
\begin{algorithm}[!ht]
\caption{The functional Riemannian Gauss--Newton scheme for minimizing \eqref{eq:loss}}
\label{alg:est}
\KwIn{response vector $\bm{y}\in\mathbb{R}^{n}$, weighted covariate tensor $\bm{\mathcal{Z}}\in\mathbb{R}^{n\times p_{0}\times p_{1} \times\dots\times p_{D}}$, Tucker rank $\bm{r} = (r_{0},r_{1},\dots,r_{D})$, multiple of penalty matrix $\rho \bm{A}_{m}$, iteration number $K$}
\Begin{
    Initialize $\bm{\varTheta}^{0} = \bm{\mathcal{S}}^{0}\times_{d=0}^{D}\bm{U}_{d}^{0}$\;
    \For{$k = 0,1,\ldots,K-1$}{
   	\For{$d = 0,1,\ldots,D$}{
            Calculate $\bm{V}_{d}^{k} = \operatorname{QR}\{\mathscr{M}_{d}(\bm{\mathcal{S}}^{k})^{\top}\}$ and $\bm{W}_{d}^{k}$ as in \eqref{eq:row_space_matrix}\;
  	}
		Obtain $\check{\bm{\varTheta}}^{k+1}$ using \eqref{eq:update_half}\;
		Update $\bm{\varTheta}^{k+1} = \bm{\mathcal{S}}^{k+1}\times_{d=0}^{D}\bm{U}_{d}^{k+1} = \mathcal{H}_{\bm{r}}(\check{\bm{\varTheta}}^{k+1})$ where $\mathcal{H}_{\bm{r}}$ is T-HOSVD\;
    }
}
\KwOut{$\widehat{\bm{\varTheta}} = \bm{\varTheta}^{K}$}
\end{algorithm}

\section{Theoretical Results}
\label{sec:theoretical}
In this section, we analyze the error bound of the functional Riemannian Gauss--Newton algorithm in Subsection \ref{subsec:RGN}. It will be shown that the convergence rate is quadratic which outperforms the commonly used first-order methods. 
{We consider varying samples sizes $n$ and dimensions $p_d, r_d$, $d=0,1,\dots,D$. 

Based on \eqref{eq:KL_decomp} and \eqref{eq:obs}, we require the following assumptions to guarantee the efficiency of estimation. 
\begin{assumption}
\label{ass:eigen_val}
For each $k$, the tensor $\bm{\varXi}_{k}$ has {entries with zero mean. Moreover, there exist some constants $1<a,A<\infty$ such that \[ A^{-1}k^{-a}\norm{\bm{\varUpsilon}}_{\mathrm{F}}^{2} \le E(\langle\bm{\varXi}_{k},\bm{\varUpsilon}\rangle^{2}) \le Ak^{-a}\norm{\bm{\varUpsilon}}_{\mathrm{F}}^{2} ,\quad \bm{\varUpsilon}\in\mathbb{R}^{p_1\times\dots\times p_D} .\]} 
\end{assumption}
\begin{assumption}
\label{ass:eigen_fn_ortho}
Let $\bm{\Phi}_{\ell}$ be the $p_{0}$-by-$p_{0}$ matrix with \[ [\bm{\Phi}_{\ell}]_{j,k} = (\Delta t_{j})^{1/2} \varphi_{k+(\ell-1)p_{0}}(t_{j}) ,\quad j,k=1,\ldots,p_{0} .\] There exists some constant $0<C_{\varphi}<\infty$ such that for $\ell = 1,2,\ldots$, the spectral norm \[ \sigma_{1}(\bm{\Phi}_{\ell}) \le C_{\varphi} .\]
\end{assumption}
\begin{assumption}
\label{ass:obs_noise}
The random tensors $\bm{\mathcal{E}}_{ij}$ in \eqref{eq:obs} are independent and have uncorrelated entries with mean zero and variance $\sigma_{X}^{2} < \infty$.
\end{assumption}

Assumption~\ref{ass:eigen_val} imposes smoothness in terms of the decay of {covariance operators of $\bm{\varXi}_{k}$}, which in dimension one coincides with the smoothness condition stated in the literature on functional data analysis \citep{cai2006prediction,hall2007methodology}. Assumption~\ref{ass:eigen_fn_ortho} implies that the eigenfunctions capture signals {with bounded transformations} and prevent the functional pattern from being too wild, {that is, the weighted sampling matrices $\bm{\Phi}_{\ell}$ have bounded operator norms}. Indeed, $\bm{\Phi}_{\ell}$'s turn out to be orthogonal matrices when, for instance, $\varphi_{k}(t) = \sin(2k\pi t)$ and $t_{j} = (j-1/2)/p_{0}$. Then as illustrated by \eqref{eq:TRIP_upper} and \eqref{eq:TRIP_lower} in Lemma \ref{lem:TRIP}, we are able to offer a substitute for the technical tensor restricted isometry property in \citet{luo2023low}; see also Remark \ref{rmk:size} for the requirement of the sample size $n$ when the considered tensors admit low Tucker rank.
\begin{lemma}\label{lem:TRIP}
Suppose that Assumptions \ref{ass:eigen_val}--\ref{ass:obs_noise} hold. 
For any $\bm{\varUpsilon}\in\mathbb{R}^{p_{0}\times p_{1}\times \dots\times p_{D}}$, the probability that 
\begin{equation}\label{eq:TRIP_upper}
n^{-1} \norm{\mathscr{Z}\bm{\varUpsilon}}^{2} \le R_{u} p_{0}^{-1} \norm{\bm{\varUpsilon}}_{\mathrm{F}}^{2}
\end{equation}
tends to one as $n\to\infty$, where $R_{u} = (a-1)^{-1}(2a-1)AC_{0}C_{\varphi}+C_{0}^{2}\sigma_{X}^{2}$. 
Furthermore, writing $\bm{\Phi}_{0}$ for the $p_{0}$-by-$r_{0}$ matrix such that $[\bm{\Phi}_{0}]_{j,k} = (\Delta t_{j})^{1/2} \varphi_{k}(t_{j})$, if for some $c>0$, 
\begin{equation}\label{eq:eigen_trunc}
\lVert\bm{\varUpsilon}\times_{0}\bm{\Phi}_{0}^{\top}\rVert_{\mathrm{F}} \ge c \norm{\bm{\varUpsilon}}_{\mathrm{F}} , 
\end{equation}
then the probability that 
\begin{equation}\label{eq:TRIP_lower}
n^{-1} \norm{\mathscr{Z}\bm{\varUpsilon}}^{2} \ge R_{l} p_{0}^{-1} \norm{\bm{\varUpsilon}}_{\mathrm{F}}^{2}
\end{equation}
tends to one as $n\to\infty$, where $R_{l} = c^{2}/(AC_{0}r_{0}^{a})$. 
\end{lemma}
{Compared with the tensor restricted isometry property, in Lemma~\ref{lem:TRIP} we relax the requirement that $R_l \in (0,1)$ and $R_u \in (1,2)$.}
Condition \eqref{eq:eigen_trunc} means that the functional mode of $\bm{\varUpsilon}$ is well accounted for by the leading $r_{0}$ eigenfunctions, which is reasonable in the estimation procedure. We make this more precise in the following lemma.
\begin{lemma}\label{lem:trunc}
For $\bm{\varUpsilon}\in\mathbb{R}^{p_{0}\times p_{1}\times \dots\times p_{D}}$ and $\bm{\Phi}\in\mathbb{R}^{p_{0}\times r_{0}}$, it holds that 
\[ \lVert\bm{\varUpsilon}\times_{0}\bm{\Phi}^{\top}\rVert_{\mathrm{F}} \ge \big\{ \sigma_{r_{0}}(\bm{\Phi}^{\top}\bm{U}_{0})\alpha - \sigma_{1}(\bm{\Phi}^{\top}\bm{U}_{0\perp})(1-\alpha^{2})^{1/2} \big\} \norm{\bm{\varUpsilon}}_{\mathrm{F}} , \]
where $\bm{U}_{0}\in\mathbb{O}_{p_{0},r_{0}}$ consists of the leading $r_{0}$ columns of $\operatorname{SVD}\{\mathscr{M}_{0}(\bm{\varUpsilon})\}$, and $\alpha\in(0,1]$ such that $\alpha\norm{\bm{\varUpsilon}}_{\mathrm{F}} = \norm{\bm{\varUpsilon}\times_{0}(\bm{U}_{0}\bm{U}_{0}^{\top})}_{\mathrm{F}}$.
\end{lemma}

With the help of Lemma \ref{lem:TRIP}, we can establish the following deterministic convergence theory.
Let 
\begin{equation}\label{eq:coef_norm}
C_{m} = \int_{\mathbb{T}} \big( \norm{\bm{\mathcal{B}}(t)}_{\mathrm{F}}^{2} + \lVert\bm{\mathcal{B}}^{(m)}(t)\rVert_{\mathrm{F}}^{2} \big) \dd{t} 
\end{equation}
and 
\begin{equation}\label{eq:warm}
\zeta_{m} = \sup\Big\{ \zeta\ge 0 : \sup_{\varUpsilon\in\mathbb{M}_{\bm{r}} : \norm{\bm{\varUpsilon}-\bm{\varTheta}}_{\mathrm{F}} \le \zeta} \langle \mathscr{A}\mathscr{P}_{\bm{\varUpsilon}}\bm{\varTheta} , \mathscr{P}_{\bm{\varUpsilon}}\bm{\varTheta} \rangle \le 2 C_{m} \Big\} .
\end{equation}
\begin{theorem}\label{thm:rate}
Recall the updates \eqref{eq:update_half} and \eqref{eq:update} in the functional Riemannian Gauss--Newton algorithm for minimizing \eqref{eq:loss}.
If \eqref{eq:TRIP_upper} holds for $\bm{\varUpsilon} = \bm{\varTheta}-\mathscr{P}_{\bm{\varTheta}^{k}}\bm{\varTheta}$, \eqref{eq:TRIP_lower} holds for $\bm{\varUpsilon} = \check{\bm{\varTheta}}^{k+1}-\bm{\varTheta}$, and $\norm{\bm{\varTheta}^{k}-\bm{\varTheta}}_{\mathrm{F}} \le \zeta_{m}$, 
then 
\begin{equation}\label{eq:rate}
\begin{aligned}
\lVert\bm{\varTheta}^{k+1}-\bm{\varTheta}\rVert_{\mathrm{F}} \le {}& \{(D+1)^{1/2}+1\}(D+1)8^{1/2}R_{u}^{1/2}R_{l}^{-1/2}\lambda_{\min}^{-1} \lVert\bm{\varTheta}^{k}-\bm{\varTheta}\rVert_{\mathrm{F}}^{2} \\ & + \{(D+1)^{1/2}+1\}R_{l}^{-1/2}\eta , 
\end{aligned}
\end{equation}
where $\lambda_{\min} = \min_{0\le d\le D} \sigma_{r_{d}}\{\mathscr{M}_{d}(\bm{\varTheta})\}$, and $\eta>0$ is defined by 
\begin{equation}\label{eq:opt_err}
\eta^{2} = 12p_{0}n^{-1}\norm{\bm{\delta}}^{2} + 16\rho p_{0}C_{m} + 2c_{m}^{-1}p_{0}^{2}n^{-2}\rho^{-1} \norm{(\mathscr{Z}^{*}\bm{\varepsilon})_{\max(2\bm{r})}}_{\mathrm{F}}^{2} .
\end{equation}
Here $\bm{\delta}$ is the vector of approximation errors \[ \delta_{i} = \int_{\mathbb{T}} \langle \bm{\mathcal{X}}_{i}(t) , \bm{\mathcal{B}}(t) \rangle \dd{t} - \sum_{j=1}^{p_0} \langle \bm{\mathcal{X}}_{i}(t_{j}) , \bm{\mathcal{B}}_{j} \rangle \Delta t_{j} ,\quad i=1,\ldots,n ,\] and $c_{m}$ is given in \eqref{eq:pen_low_bound}.
\end{theorem}

The error bound \eqref{eq:rate} 
ensures the functional Riemannian Gauss--Newton iterate to converge quadratically to the ball centered at $\bm{\varTheta}$ of radius $\order{\eta}$. The additional term $\eta^{2}$ includes three parts that result from approximation (first summand), regularization (second), and observational noise (third), respectively. We defer the investigation of $\eta$ to Theorem \ref{thm:err}. Looking closer at the right-hand side of \eqref{eq:rate}, one can partition the process of convergence into two phases. 
\begin{corollary}\label{cor:phase}
Under the assumptions in Theorem \ref{thm:rate}, 

if $\lVert\bm{\varTheta}^{k}-\bm{\varTheta}\rVert_{\mathrm{F}} \ge 8^{-1/4}R_{u}^{-1/4}(D+1)^{-1/2}\lambda_{\min}^{1/2}\eta^{1/2}$, then \[ \lVert\bm{\varTheta}^{k+1}-\bm{\varTheta}\rVert_{\mathrm{F}} \le 2\{(D+1)^{1/2}+1\}(D+1)8^{1/2}R_{u}^{1/2}R_{l}^{-1/2}\lambda_{\min}^{-1} \lVert\bm{\varTheta}^{k}-\bm{\varTheta}\rVert_{\mathrm{F}}^{2} ,\] which renders quadratic convergence; and

if $\lVert\bm{\varTheta}^{k}-\bm{\varTheta}\rVert_{\mathrm{F}} \le 8^{-1/4}R_{u}^{-1/4}(D+1)^{-1/2}\lambda_{\min}^{1/2}\eta^{1/2}$, then \[ \lVert\bm{\varTheta}^{k+1}-\bm{\varTheta}\rVert_{\mathrm{F}} \le 2\{(D+1)^{1/2}+1\}R_{l}^{-1/2}\eta ,\] which reflects the eventual estimation error. 

Consequently, if $\norm{\bm{\varTheta}^{0}-\bm{\varTheta}}_{\mathrm{F}} \le 4^{-1}\{(D+1)^{1/2}+1\}^{-1}(D+1)^{-1}8^{-1/2}R_{u}^{-1/2}R_{l}^{1/2}\lambda_{\min}$, then 
\[ \lVert\bm{\varTheta}^{k}-\bm{\varTheta}\rVert_{\mathrm{F}} \le 2^{-2^{k}}\lVert\bm{\varTheta}^{0}-\bm{\varTheta}\rVert_{\mathrm{F}} \]
for $k \le K = \inf\{ k : \norm{\bm{\varTheta}^{k}-\bm{\varTheta}}_{\mathrm{F}} \le 8^{-1/4}R_{u}^{-1/4}(D+1)^{-1/2}\lambda_{\min}^{1/2}\eta^{1/2} \}$; 
and 

if $8^{-1/4}R_{u}^{-1/4}(D+1)^{-1/2}\lambda_{\min}^{1/2}\eta^{1/2} \ge 2\{(D+1)^{1/2}+1\}R_{l}^{-1/2}\eta$, then 
\[ \lVert\bm{\varTheta}^{k}-\bm{\varTheta}\rVert_{\mathrm{F}} \le 2\{(D+1)^{1/2}+1\}R_{l}^{-1/2}\eta \]
for $k > K$. The above-defined $K$ satisfies that 
\[ K \le \lceil \log\max\{1, 2^{-1}\log 16^{-1}\{(D+1)^{1/2}+1\}^{-2}(D+1)^{-1}8^{-1/2}R_{u}^{-1/2}R_{l}\lambda_{\min}\eta^{-1} \} \rceil .\]
\end{corollary}

Corollary \ref{cor:phase} implies that we only need $\order{\log\log\eta^{-1}}$ iterations to achieve the estimation error bound. 
The magnitude of $\eta$ plays an essential role in both the estimation error bound and the required number of iterations, and we show it in Theorem \ref{thm:err}. To characterize the approximation error, we invoke the concept of H\"{o}lder continuity. A function $\bm{\mathcal{X}} : \mathbb{T} \to \mathbb{R}^{p_{1}\times\dots\times p_{D}}$ is said to be $\kappa$-H\"{o}lder continuous if $\norm{\bm{\mathcal{X}}(t)-\bm{\mathcal{X}}(s)}_{\mathrm{F}}/\abs{t-s}^{\kappa}$ is uniformly bounded for $t,s \in \mathbb{T}$ with $t\ne s$.
\begin{theorem}\label{thm:err}
Suppose that Assumptions \ref{ass:eigen_val}--\ref{ass:obs_noise} hold and that $\bm{\mathcal{X}}(\bm{\cdot})$ is almost surely $\kappa$-H\"{o}lder continuous for some $0<\kappa\le1$. If $\varepsilon$ follows the normal distribution $\mathcal{N}(0,\sigma_{y}^{2})$, then $\eta$ defined by \eqref{eq:opt_err} satisfies that 
\begin{equation*}
\eta^{2} = \mathcal{O}_{\mathrm{pr}}\bigg\{ p_{0}^{1-2\kappa} C_{m} + \rho p_{0} C_{m} + (n\rho)^{-1}p_{0}\Big( \sum_{d=0}^{D}p_{d}r_{d} + \prod_{d=0}^{D}r_{d} \Big) \bigg\} .
\end{equation*}
If $\rho\asymp\big\{n^{-1}\big(\sum_{d=0}^{D}p_{d}r_{d} + \prod_{d=0}^{D}r_{d}\big)/C_{m}\big\}^{1/2}$, then the bound can be reduced to
\begin{equation}\label{eq:err_rate_final}
\eta^{2} = \mathcal{O}_{\mathrm{pr}}\bigg\{ p_{0}^{1-2\kappa} C_{m} + n^{-1/2}p_{0} \Big( \sum_{d=0}^{D}p_{d}r_{d} + \prod_{d=0}^{D}r_{d} \Big)^{1/2} C_{m}^{1/2} \bigg\} . 
\end{equation}
\end{theorem}
{The normality of $\varepsilon$ is required in Theorem~\ref{thm:err} so that $\norm{(\mathscr{Z}^{*}\bm{\varepsilon})_{\max(2\bm{r})}}_{\mathrm{F}}$ has a light-tailed distribution, given Lemma~\ref{lem:TRIP} that controls the operator norm of $\mathscr{Z}$.}

Due to \eqref{eq:interp}, the quantity $C_{m}$ defined in \eqref{eq:coef_norm} has the same order with $p_{0}^{-1} \norm{\bm{\varTheta}}_{\mathrm{F}}^{2}$. By \citet[Lemma 12]{luo2023low}, \[ \norm{\bm{\varTheta}}_{\mathrm{F}}^{2} = \mathcal{O}_{\mathrm{pr}}\Big( \sum_{d=0}^{D}p_{d}r_{d} + \prod_{d=0}^{D}r_{d} \Big) \] if $\bm{\varTheta} = \bm{\varUpsilon}_{\max(\bm{r})}$ for some $\bm{\varUpsilon}\in\mathbb{R}^{p_{0}\times p_{1}\times\dots\times p_{D}}$ with i.i.d.\ $\mathcal{N}(0,1)$ entries. In such a case, combining Corollary \ref{cor:phase} and Theorem \ref{thm:err}, we conclude that the final estimator $\widehat{\bm{\varTheta}}$ has the following error upper bound 
\begin{equation}\label{eq:bound_Theta}
\norm{\widehat{\bm{\varTheta}}-\bm{\varTheta}}_{\mathrm{F}}^{2} = \order{\eta^{2}} = \mathcal{O}_{\mathrm{pr}}\bigg\{ \big( p_{0}^{-2\kappa} + n^{-1/2}p_{0}^{1/2} \big) \Big( \sum_{d=0}^{D}p_{d}r_{d} + \prod_{d=0}^{D}r_{d} \Big) \bigg\}
\end{equation}
when the tuning parameter $\rho\asymp(p_{0}/n)^{1/2}$. 
Based on \eqref{eq:interp}, let $\widehat{\bm{\mathcal{B}}}(\bm{\cdot})$ be the plug-in estimator using $\widehat{\bm{\varTheta}}$. It is clear that \[ \int_{\mathbb{T}} \norm{\widehat{\bm{\mathcal{B}}}(t)-\bm{\mathcal{B}}(t)}_{\mathrm{F}}^{2} \dd{t} = \left\langle \big(\widehat{\bm{\varTheta}}-\bm{\varTheta}\big) \times_{0} \bm{\Omega}_{0}, \widehat{\bm{\varTheta}}-\bm{\varTheta} \right\rangle ,\] where $\bm{\Omega}_{0}$ is defined similarly as \eqref{eq:penalty-der}. Since $\bm{\Omega}_{0} \asymp p_{0}^{-1} \bm{I}_{p_0}$ as $p_{0}\to\infty$, we deduce from \eqref{eq:bound_Theta} that 
\begin{equation*}
\int_{\mathbb{T}} \norm{\widehat{\bm{\mathcal{B}}}(t)-\bm{\mathcal{B}}(t)}_{\mathrm{F}}^{2} \dd{t} = \mathcal{O}_{\mathrm{pr}}\bigg[ \big\{ p_{0}^{-2\kappa-1} + (np_{0})^{-1/2} \big\} \Big( \sum_{d=0}^{D}p_{d}r_{d} + \prod_{d=0}^{D}r_{d} \Big) \bigg].
\end{equation*}
{Furthermore, given $\int_{\mathbb{T}} \norm{\bm{\mathcal{B}}(t)}_{\mathrm{F}}^{2} \dd{t} \asymp p_{0}^{-1} \norm{\bm{\varTheta}}_{\mathrm{F}}^{2} \asymp p_{0}^{-1}\big( \sum_{d=0}^{D}p_{d}r_{d} + \prod_{d=0}^{D}r_{d} \big)$, it is seen that 
\begin{equation}\label{eq:RISE}
\int_{\mathbb{T}} \norm{\widehat{\bm{\mathcal{B}}}(t)-\bm{\mathcal{B}}(t)}_{\mathrm{F}}^{2} \dd{t} \bigg/ \int_{\mathbb{T}} \lVert\bm{\mathcal{B}}(t)\rVert_{\mathrm{F}}^{2} \dd{t} = \mathcal{O}_{\mathrm{pr}}\big\{ p_{0}^{-2\kappa} + n^{-1/2}p_{0}^{1/2} \big\} .
\end{equation}
The first term decreasing with $p_0$ coincides with the traditional wisdom of nonparametric smoothing, while the second term increasing with $p_0$ reflects the complicated dimensionality in the tensor structure.
It is suspected that the modeling of $\bm{\mathcal{B}}(\cdot)$ in \eqref{eq:interp} may need more smoothness, and a possible direction is considering $\bm{\mathcal{B}}(\cdot)$ to be a smooth map valued in a low-rank tensor space. We leave this as a potential future work.}

In the following theorem, we demonstrate that the bound \eqref{eq:bound_Theta} is optimal in a minimax sense with respect to the tensor dimensionality. {We do not tackle the functional aspect due to its complexity, which, as seen from \eqref{eq:RISE}, differs from the classical rate of convergence for nonparametric estimators.}
\begin{theorem}\label{thm:minimax}
Consider the model $\mathfrak{M}$ of $(\bm{y},\mathscr{Z},\bm{\varTheta})$ such that $\mathscr{Z}$ satisfies \eqref{eq:TRIP_upper} and \eqref{eq:TRIP_lower} for $\bm{\varUpsilon}\in\bigcup_{\bm{s}\le2\bm{r}}\mathbb{M}_{\bm{s}}$ and that $\bm{\varepsilon}=\bm{y}-\mathscr{Z}\bm{\varTheta}$ satisfies $\norm{\bm{\varepsilon}}^{2} = \mathcal{O}_{\mathrm{pr}}(\xi)$ with some $\xi \ge \sum_{d=0}^{D}p_{d}r_{d}+\prod_{d=0}^{D}r_{d}$. Then any estimator $\widetilde{\bm{\varTheta}}$ for $\bm{\varTheta}\in\mathbb{M}_{\bm{r}}$ based on $\bm{y}$ and $\mathscr{Z}$ satisfies that 
\[ \sup_{n\le\prod_{d=0}^{D}p_{d}} n^{1/2}p_{0}^{-1/2} \sup_{(\bm{y},\mathscr{Z},\bm{\varTheta})\in\mathfrak{M}} \lVert\widetilde{\bm{\varTheta}}-\bm{\varTheta}\rVert_{\mathrm{F}} \ge 2^{-1/2} \norm{\bm{\mathcal{E}}_{0}}_{\mathrm{F}} \]
for some random tensor $\bm{\mathcal{E}}_{0}$ such that \[ E(\norm{\bm{\mathcal{E}}_{0}}_{\mathrm{F}}^{2}) \ge c_{0}\Big( \sum_{d=0}^{D}p_{d}r_{d}+\prod_{d=0}^{D}r_{d} \Big) \] where $c_{0}>0$ is a constant depending only on $D$.
\end{theorem}

As far as the difficulty of tensor computation is concerned, we end this section with the following remarks. {Note that efficient tensor methods often turn multiplicative costs into additive.}
\begin{remark}[Initialization]
In Theorem~\ref{thm:rate}, the quantity $\zeta_{m}$ in \eqref{eq:warm} is well defined since $\lim_{\bm{\varUpsilon}\to{\varTheta}}\mathscr{P}_{\bm{\varUpsilon}}\bm{\varTheta}=\bm{\varTheta}$ \citep[Lemma 9]{luo2023low}. 
{Theorem~\ref{thm:rate} and Corollary~\ref{cor:phase} require a suitable initialization, namely, $\norm{\bm{\varTheta}^{0}-\bm{\varTheta}}_{\mathrm{F}}$ is of order $\min\{\zeta_{m}, D^{-3/2}R_{u}^{-1/2}R_{l}^{1/2}\lambda_{\min}\}$. 
The linear dependence of the bound on $\lambda_{\min}$ matches the initialization condition in the literature \citep[Remark 2]{luo2023low}, and is often satisfied when warming up the output of some decomposition algorithms \citep[Section 4]{luo2023low}. In practice, we suggest setting $\bm{\varTheta}^{0} = \mathcal{H}_{\bm{r}}(\mathscr{Z}^{*}\bm{y})$, which has worked well in our experiments.} 
\end{remark}
\begin{remark}[Sample size]\label{rmk:size}
In view of \eqref{eq:err_rate_final}, for estimation accuracy we need the sample size $n$ to exceed $\mathcal{O}\{ p_{0}^{2}( \sum_{d=0}^{D}p_{d}r_{d} + \prod_{d=0}^{D}r_{d} )C_{m} \}$. Note also that in order to fulfill \eqref{eq:TRIP_lower}, \citet[Theorem 2]{rauhut2017low} suggests that $n = \mathcal{O}( \sum_{d=0}^{D}p_{d}r_{d}+\prod_{d=0}^{D}r_{d} )$ could be adequate if $\rank_{\mathrm{Tuc}}(\bm{\varUpsilon}) = (r_{0},r_{1},\dots,r_{D})$ and the entries of $\bm{\varXi}_{k}$ are sub-Gaussian for $k=1,\ldots,r_{0}$. 
\end{remark}
\begin{remark}[Computational complexity]
In each iteration of the functional Riemannian Gauss--Newton algorithm, the computational cost for obtaining the update \eqref{eq:update} is acceptable. Denote $\bar{p} = \max_{d} p_{d}$ and $\bar{r} = \max_{d} r_{d}$. Computing $\mathscr{R}_{\bm{\varTheta}^{k}}^{*}\mathscr{Z}^{*}$ involves $\mathscr{R}_{\bm{\varTheta}^{k}}^{*}\bm{\mathcal{Z}}_{i}$ for $i=1,\ldots,n$, so the number of operations is $\mathcal{O}(n\bar{p}^{D+1}\bar{r})$. By Proposition \ref{prop:LAL}, computing $\mathscr{R}_{\bm{\varTheta}^{k}}^{*}\mathscr{A}\mathscr{R}_{\bm{\varTheta}^{k}}$ only needs $\mathcal{O}(\bar{p}^{2}\bar{r})$ operations. Then it takes $\mathcal{O}[ n\{\bar{r}^{D+1}+(D+1)\bar{p}\bar{r}\}^{2} ]$ operations to solve the linear equation system in $\mathbb{D}$. Finally, extending the solution to $\mathrm{T}_{\bm{\varTheta}^{k}}\mathbb{M}_{\bm{r}}$ uses $\mathcal{O}(\bar{p}^{D+1}\bar{r})$ operations. In summary, the cost is $\mathcal{O}(n\bar{p}^{D+1}\bar{r})$, provided that $\bar{r}=o(\bar{p}^{1/2})$. 
\end{remark}

\section{Numerical Studies}
\label{sec:numerical}
We perform the proposed method on simulation and real data. In this section, the numerical results are reported. 

\subsection{Simulation}
We generate a sample of $n=500$ subjects with functional tensor covariates of order $(p_{0},p_{1},p_{2})=(12,8,8)$. The functional tensors are $\bm{\mathcal{X}}_{i}(t) = \sum_{k=1}^{30} k^{-1} \sin(k\pi t) \bm{\varUpsilon}_{ik} ,\ i=1,\ldots,n$, observed as $\bm{\mathcal{X}}_{ij}$ at the grid points $t_{j}=(j-1/2)/p_{0} ,\ j=1,\ldots,p_{0}$. Here $\bm{\varUpsilon}_{ik}$'s are i.i.d.\ random tensors with i.i.d.\ $\mathcal{N}(0,1)$ entries and the noise $\bm{\mathcal{E}}_{ij} = \bm{\mathcal{X}}_{ij} - \bm{\mathcal{X}}_{i}(t_{j})$ are i.i.d.\ random tensors with i.i.d.\ $\mathcal{N}(0,0.05^{2})$ entries. The regression coefficient in \eqref{eq:interp} is generated by $\bm{\varTheta} = \bm{\mathcal{S}} \times_{d=0}^{2} \bm{U}_{d}$ where $\bm{\mathcal{S}}$ is a random tensor of order $(r_{0},r_{1},r_{2})=(2,3,3)$ with i.i.d.\ $\mathcal{N}(0,1)$ entries and $\bm{U}_{d}$'s are independent and uniformly distributed over $\mathbb{O}_{p_{d},r_{d}}$, and cubic natural splines are used. Each response $y_{i}$ is associated with an observational error $\varepsilon_{i}$ from $\mathcal{N}(0,0.1^{2})$. The initialization for the functional Riemannian Gauss--Newton algorithm is chosen to be $\bm{\varTheta}^{0} = \mathcal{H}_{\bm{r}}(\mathscr{Z}^{*}\bm{y})$. Given a tuning parameter $\rho$, we terminate the algorithm when the number of iterations reaches $80$ or the relative error $\norm{\bm{\varTheta}^{k}-\bm{\varTheta}}_{\mathrm{F}}/\norm{\bm{\varTheta}}_{\mathrm{F}}$ is less than $10^{-8}$. 
The first panel of Figure~\ref{fig:simu} illustrates the quadratic convergence of our proposed algorithm in several Monte Carlo replications. Due to the observational error, the obtained estimates lie in a neighborhood of the true parameter. 
{The adoption of a uniform grid could be relaxed, and we present an example of non-uniform grid points in the supplementary material, which achieves performance similar to the uniform case.}

To select the tuning parameter $\rho$, we minimize the generalized cross-validation criterion, 
\[ \mathrm{GCV} = \frac{n^{-1}\norm{\bm{y}-\bm{H}_{\rho}\bm{y}}^{2}}{\{1-n^{-1}\tr(\bm{H}_{\rho})\}^{2}} ,\]
where $\bm{H}_{\rho} = n^{-1} \mathscr{Z}\mathscr{R}_{\widehat{\bm{\varTheta}}} \big( n^{-1}\mathscr{R}_{\widehat{\bm{\varTheta}}}^{*}\mathscr{Z}^{*}\mathscr{Z}\mathscr{R}_{\widehat{\bm{\varTheta}}} + \rho \mathscr{R}_{\widehat{\bm{\varTheta}}}^{*}\mathscr{A}\mathscr{R}_{\widehat{\bm{\varTheta}}} \big)^{-1} \mathscr{R}_{\widehat{\bm{\varTheta}}}^{*}\mathscr{Z}^{*}$. 
The quality of our estimator is assessed by the relative integrated squared error, 
\[ \mathrm{RISE} = \int_{0}^{1} \norm{\widehat{\bm{\mathcal{B}}}(t)-\bm{\mathcal{B}}(t)}_{\mathrm{F}}^{2} \dd{t} \bigg/ \int_{0}^{1} \lVert\bm{\mathcal{B}}(t)\rVert_{\mathrm{F}}^{2} \dd{t} ,\]
with $\widehat{\bm{\mathcal{B}}}(\bm{\cdot})$ being the plug-in estimator using the output $\widehat{\bm{\varTheta}}$ of the algorithm. Under different choices of $\rho$, the results with correct Tucker rank are presented in the last two panels of Fig.~\ref{fig:simu}. Note that $\rho=0$ corresponds to tabular tensor regression. It can be seen that the introduction of a roughness penalty improves the estimation accuracy and the selection of the tuning parameter is valid. 
{While the absolute difference of $\mathrm{RISE}$ appears small, this reflects a 1\% relative improvement. For high-dimensional data, even marginal improvements in regularization efficacy can enhance interpretability by reducing overfitting.}
To the best of our knowledge, we propose the first method for functional tensor regression. Therefore, we compare the performance of our method with standard tensor regression methods. 

\begin{figure}[!ht]
    \centering
    \includegraphics[width=0.325\linewidth]{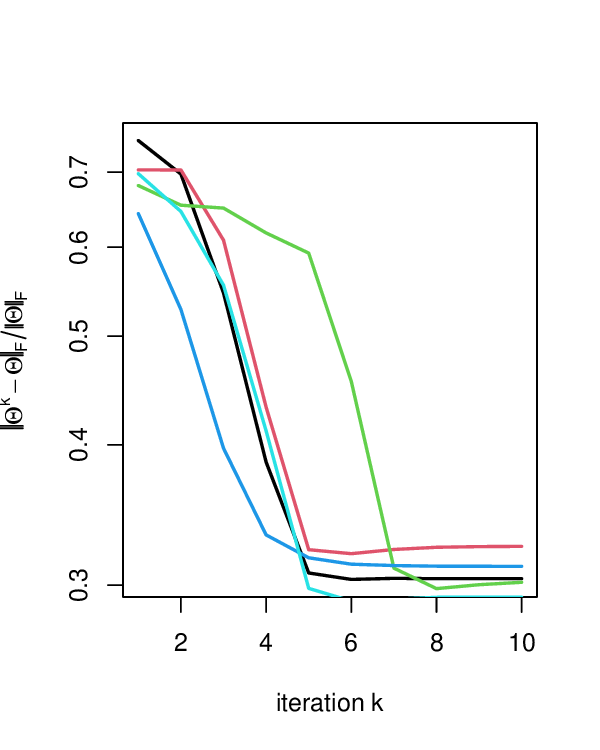}
    \includegraphics[width=0.325\linewidth]{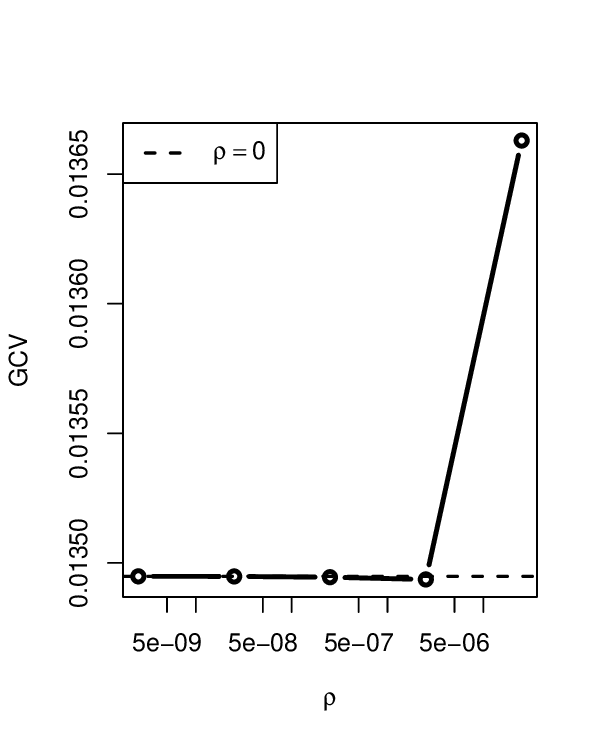}
    \includegraphics[width=0.325\linewidth]{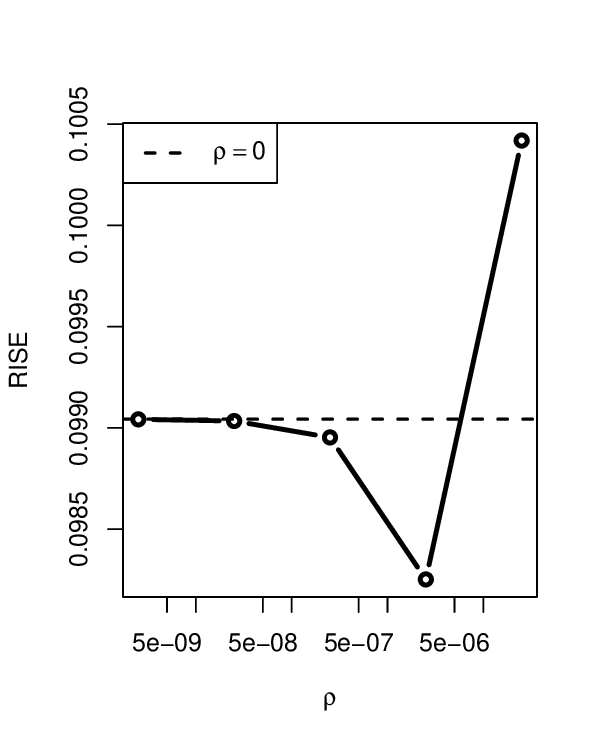}
    \caption{Left: Convergence performance of the functional Riemannian Gauss--Newton algorithm. Middle and Right: GCV and RISE versus the tuning parameter $\rho$. Displayed are averages based on 100 Monte Carlo replications of $(\bm{\mathcal{X}}_{i},y_{i})_{i=1,\ldots,500}$.}
    \label{fig:simu}
\end{figure}

We then train the functional tensor regression model using different Tucker ranks. As shown in Table~\ref{tbl:simu_rank}, the minimizer of GCV gives rise to correct specification of model rank and leads to fairly good estimation accuracy. 
When the number of rank candidates becomes excessively large, the shrinkage search method mentioned in \citet{spencer2022bayesian} is beneficial for reducing computation costs. 

\begin{table}[!ht]
    \centering
\caption{$\mathrm{GCV}$ and $\mathrm{RISE}$ corresponding to the selected tuning parameter under different Tucker ranks. Reported are the average and standard deviation (in the parenthesis) based on 100 Monte Carlo replications of $(\bm{\mathcal{X}}_{i},y_{i})_{i=1,\ldots,500}$. Marked with * is the true model rank.}
\label{tbl:simu_rank}
\begin{tabular}{ccc|ccc}
\hline
Used rank & $\mathrm{GCV}\times10^{2}$ & $\mathrm{RISE}\times10^{2}$ & Used rank & $\mathrm{GCV}\times10^{2}$ & $\mathrm{RISE}\times10^{2}$ \\ \hline
~(2,3,3)* & 1.35(0.01) & ~9.83(0.28) & (5,5,5) & 1.43(0.01) & 73.67(1.40) \\
 (2,4,3) & 1.36(0.01) & 10.88(0.36) & (2,3,2) & 2.09(0.02) & 30.48(0.40) \\
 (3,3,3) & 1.36(0.01) & 16.40(0.43) & (2,4,2) & 2.10(0.01) & 30.90(0.43) \\
 (3,4,3) & 1.36(0.01) & 20.80(0.49) & (3,3,2) & 2.12(0.02) & 37.11(0.59) \\
 (3,4,4) & 1.36(0.01) & 26.61(0.57) & (3,4,2) & 2.13(0.01) & 42.25(0.73) \\
 (4,4,4) & 1.37(0.01) & 37.68(0.83) & (2,2,2) & 2.53(0.02) & 44.27(0.43) \\
 (4,4,3) & 1.38(0.01) & 27.96(0.59) & (3,2,2) & 2.58(0.02) & 50.25(0.49) \\
\hline
\end{tabular}
\end{table}

To assess the efficiency of the proposed method, we further conduct simulation studies with varying amplitude of observational errors: $\varepsilon_i \sim \mathcal{N}(0, \sigma^2)$. As $\sigma$ ranges from 0.02 to 0.1, the Monte Carlo mean of $\frac{1}{n}\sum_{i=1}^{n} |y_i|$ varies from 0.348 to 0.358. The first panel of Figure~\ref{fig:simu_more} demonstrates that the RISE increases with the noise level in both functional and tabular tensor regression. However, the functional method consistently outperforms the tabular method across various signal-to-noise ratios. {Since regularization methods usually present advantage by reducing variances, the proposed estimator shows marginally better robustness at higher noise levels.}

The consistency of our estimator is illustrated in the second panel of Figure~\ref{fig:simu_more}, where we revisit the $\sigma=0.1$ scenario and incrementally adjust the sample size, documenting the corresponding changes in RISE. Notably, as expected, the RISE decreases, illustrating practical convergence.

Furthermore, we adjust the model based on the number of parameters. Specifically, with a fixed $n=500$, we vary $p_0$ in the temporal mode from 3 to 18. The results, depicted in the third panel of Figure~\ref{fig:simu_more}, show that initially, the RISE decreases due to effective regularization. Subsequently, as model complexity increases, RISE also increases. {This provides partially a numerical verification for \eqref{eq:RISE}, which incorporates both functional and tensor aspects and cannot be accounted for using only a single perspective.}

\begin{figure}[!ht]
    \centering
    \includegraphics[width=0.325\linewidth]{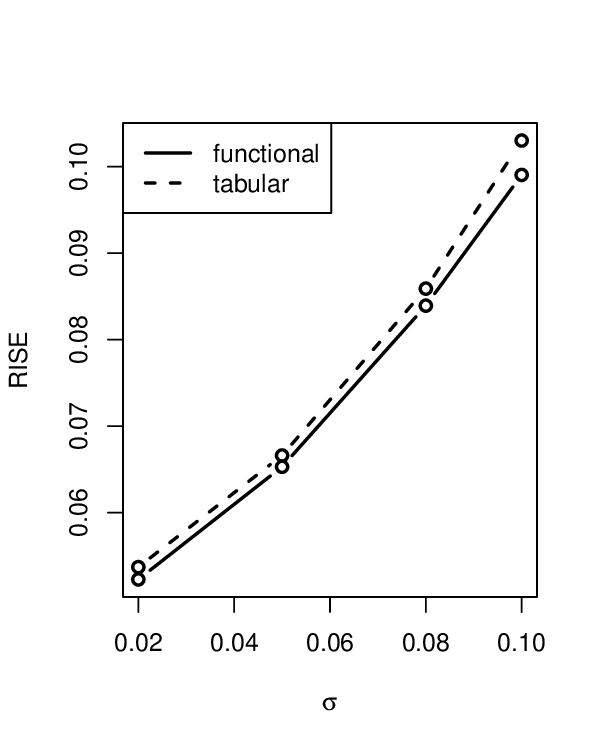}
    \includegraphics[width=0.325\linewidth]{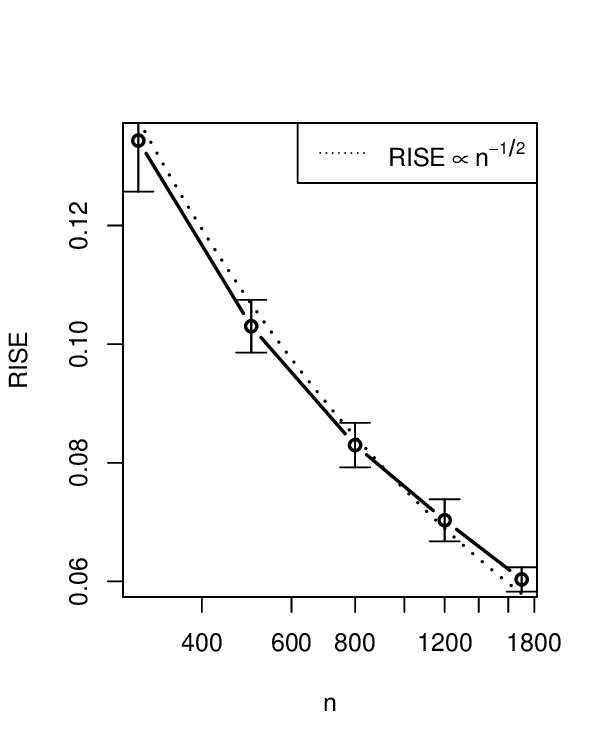}
    \includegraphics[width=0.325\linewidth]{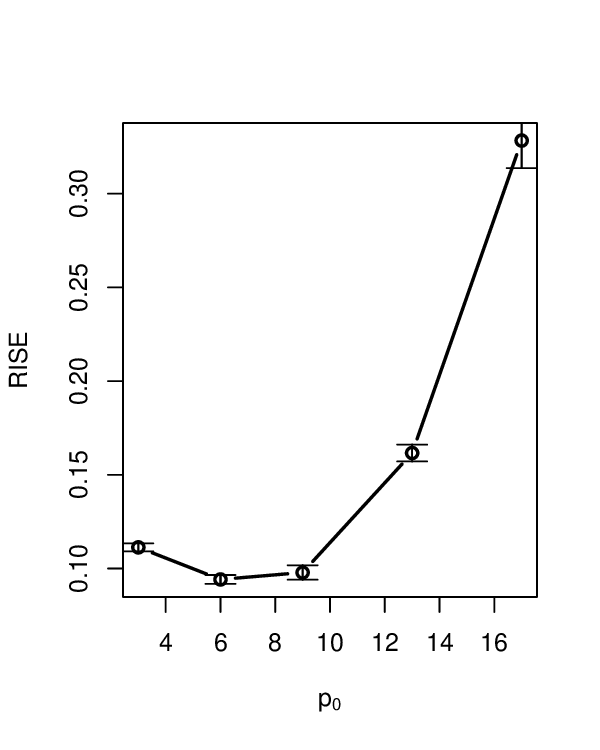}
    \caption{Left: RISE versus the standard deviation of errors. Displayed are averages based on 50 Monte Carlo replications. Solid and dashed lines correspond to functional and tabular tensor regression, respectively.
    Middle: RISE versus the sample size $n$. Solid line and error bars correspond to average and average $\pm$ standard deviation, respectively, all based on 50 Monte Carlo replications of $(\bm{\mathcal{X}}_{i},y_{i})_{i=1,\ldots,n}$. Dashed line reflects the theoretical result. The horizontal axis is in log scale to yield better visualization.
    Right: RISE versus the sampling frequency $p_0$. Displayed are average and average $\pm$ standard deviation based on 50 Monte Carlo replications.}
    \label{fig:simu_more}
\end{figure}

\subsection{Real Data Example}
Attention Deficit Hyperactivity Disorder (ADHD) is among the most common neurodevelopmental disorders of childhood. \citet{zhou2013tensor,li2018tucker} analyzed the ADHD data using tabular tensor regression models without considering the temporal effects. To remedy this, we apply our functional tensor regression model to the ADHD data. The original dataset can be downloaded from the ADHD-200 Sample Initiative (\url{http://fcon_1000.projects.nitrc.org/indi/adhd200/}), where the phenotypic test set consists of 197 subjects from 7 sites. Due to the compatibility of covariates, we extract the largest subsample from Peking University and remove 1 subject whose ADHD index is missing. Then we obtain a sample of $n=50$ subjects with ADHD index as the response. For each subject, there is an associated fMRI image that serves as the functional tensor covariate. To alleviate the computational burden, we pick $p_{0}=50$ consecutive time points, and reduce the spatial dimension to $p_{1}\times p_{2}\times p_{3} = 8\times 8\times 4$ by assuming block-wise effects. 
Five scalar covariates are also included in the regression model: gender, age, medical status, verbal IQ, and performance IQ. 

We consider the functional tensor regression model \[ y = \bm{m}^{\top}\bm{\gamma} + \int_{\mathbb{T}} \langle \bm{\mathcal{X}}(t) , \bm{\mathcal{B}}(t) \rangle \dd{t} \] where $\bm{m}$ denotes the vector of 1 and five scalar covariates. By profiling estimation using the ordinary least squares and the functional Riemannian Gauss--Newton algorithm, we fit the model with Tucker rank $(r_{0},r_{1},r_{2},r_{3}) = (2,2,2,2)$, which minimizes GCV and indicates the low-rank structure with $2$ on each of the temporal direction and the three spatial directions. Figure~\ref{fig:ADHD} depicts the estimate of $\bm{\mathcal{B}}(\bm{\cdot})$ at a series of time points on the coronal, axial, and sagittal planes, respectively, along the temporal mode from left to right and from top to bottom, and reveals some regions with lasting significant influences on the ADHD index. 
We highlight the regions with point-in-time effects, with the more significant ones distinguished by a longer active time range. 
Two such regions are cortical surfaces and white matter, studied by \cite{sowell2003cortical,makris2008attention}. Cortical surfaces refer to the outer layer of the brain, known as the cerebral cortex, whose thickness and surface area are responsible for various cognitive functions including attention, memory, and executive functioning. White matter consists of myelinated nerve fibers that facilitate communication between different regions of the brain, and thus its abnormalities like changes in the integrity of white matter tracts could bring a disorder of brain network dysfunction. 
Although the mechanism of ADHD is not fully understood due to its complexity, our results imply a potential quantification. 
To make this more illustrative, the smooth time-varying effects of the indicated regions are plotted in Figure~\ref{fig:ADHD_t}, in contrast to the cerebellum region that has little effect on ADHD index as the benchmark. The larger amplitudes corresponding to cortical surfaces and white matter reflect more important effects. 
{Besides, the sinusoidal pattern arises from the interaction between the physiological cycle and the time-varying covariate effects.}

\begin{figure}[!ht]
    \centering
\begin{minipage}{0.3\linewidth}
    \centering
    \includegraphics[width=\linewidth]{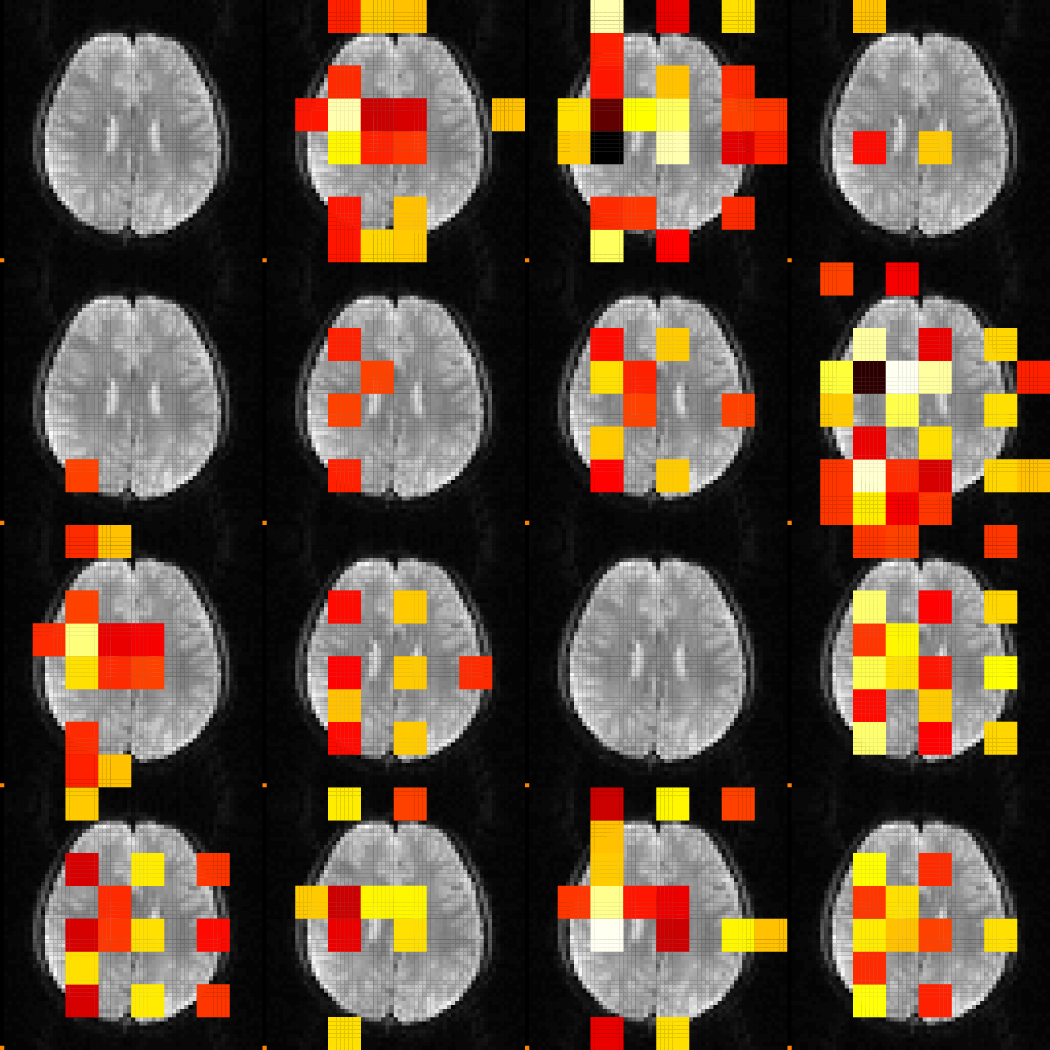}\\
    Coronal Plane 
\end{minipage}
\quad
\begin{minipage}{0.3\linewidth}
    \centering
    \includegraphics[width=\linewidth]{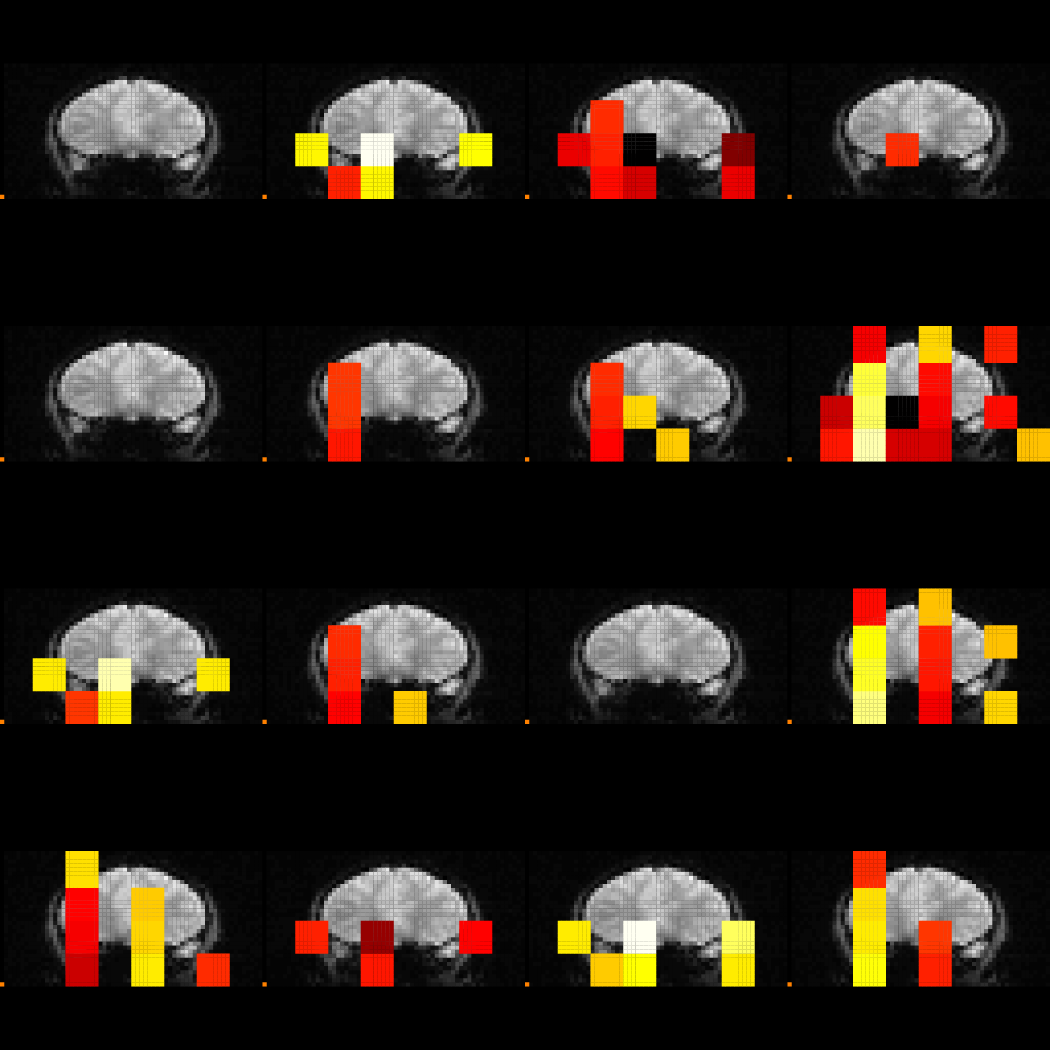}\\
    Axial Plane 
\end{minipage}
\quad
\begin{minipage}{0.3\linewidth}
    \centering
    \includegraphics[width=\linewidth]{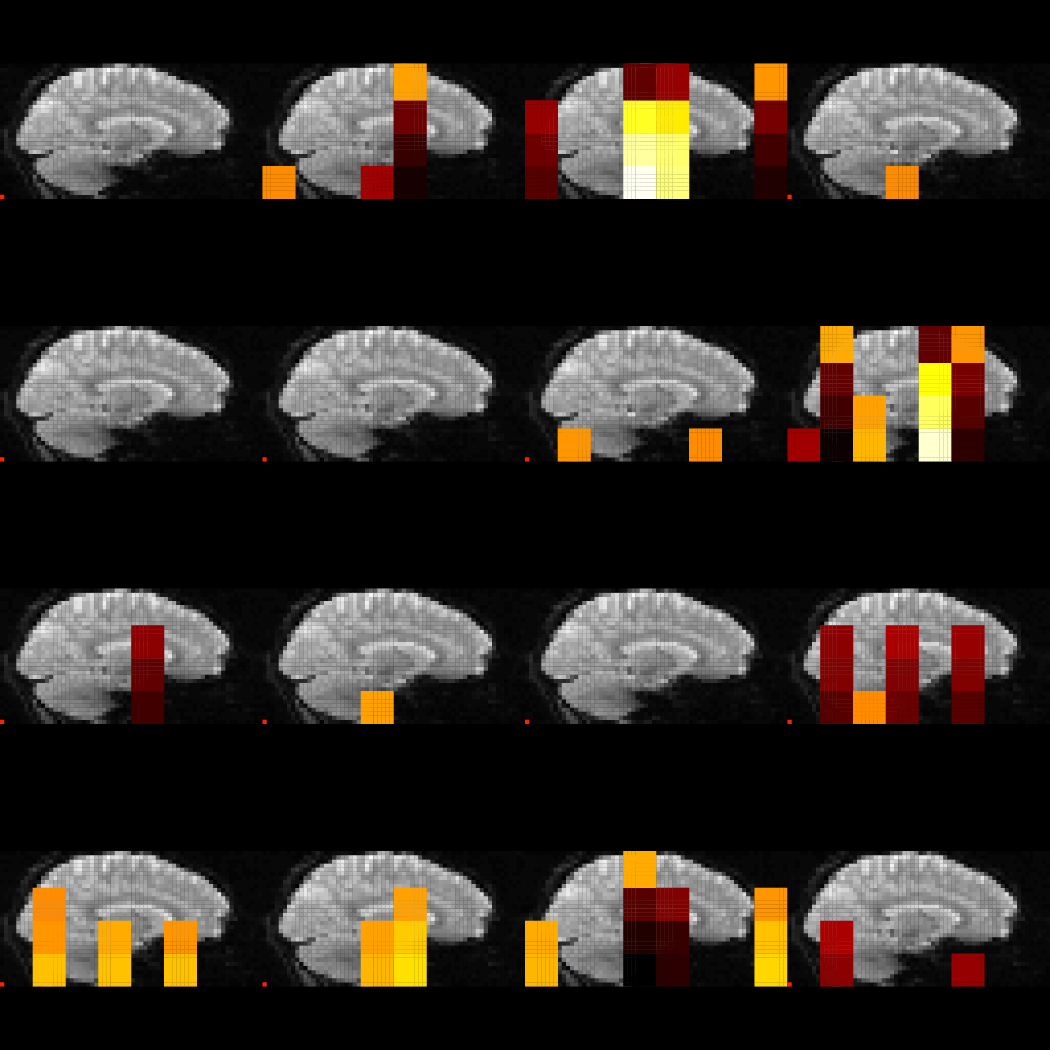}\\
    Sagittal Plane 
\end{minipage}\vspace{2ex}
    \caption{Functional tensor regression applied to the ADHD data. Plotted are slices from three spatial dimensions where only coefficients with a magnitude larger than their $80\%$ quantile are displayed. A brighter color means a larger value.}
    \label{fig:ADHD}
\end{figure}
\begin{figure}[!ht]
    \centering
    \includegraphics[width=0.7\linewidth, height=40ex]{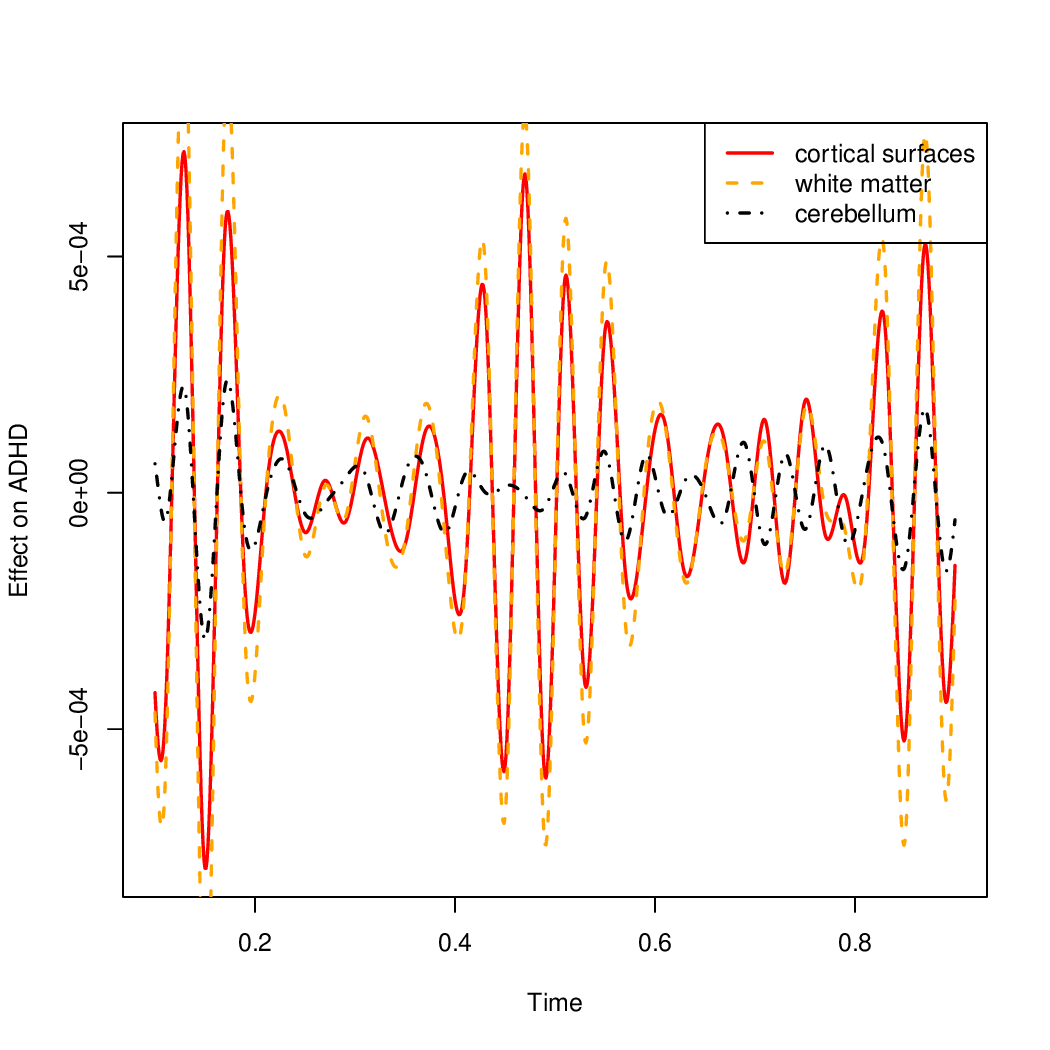}
    \caption{Estimated effects of different regions of the brain on ADHD along time.}
    \label{fig:ADHD_t}
\end{figure}

The variability of the ADHD data is well explained by our model, which can be seen from the $R$-squared. If $\bar{y}$ is the sample mean of $y_i$'s and $\hat{y}_i$ denotes the fitted value of subject~$i$, then \[ R^2 = 1 - \frac{\sum_{i=1}^{n}(y_i-\hat{y}_i)^2}{\sum_{i=1}^{n}(y_i-\bar{y})^2} = 0.991 .\]
To assess the out-of-sample prediction accuracy, we introduce the $K$-fold cross-validation, 
\[ \mathrm{CV} = \frac{1}{n}\sum_{k=1}^{K}\sum_{i\in S_k} (y_i-\hat{y}_i^{\setminus k})^2 ,\]
where the sample is split into $K$ equal-sized parts $S_1,\dots,S_K$ and $\hat{y}_i^{\setminus k}$ is the predicted value of subject~$i$ within the model trained by data from $S_1,\dots,S_{k-1},S_{k+1},\dots,S_K$. Using $K=10$, we have $\mathrm{CV}=121.9$ and $\mathrm{CV}=265.6$ for $\rho=10^{2.3}$ (selected by GCV) and $\rho=10^{-2}$, respectively. {The tabular tensor regression corresponding to $\rho=0$ is infeasible for computation due to the small sample size, so we use a sufficiently small $\rho=10^{-2}$ to offer an approximation.} This demonstrates that functional tensor regression improves upon the tabular method in the analysis of ADHD data.

\appendix

\section{T-HOSVD}
\label{sec:T-HOSVD}
Here we demonstrate the T-HOSVD operation \citep{de2000multilinear} used in \eqref{eq:update}, as shown in Algorithm~\ref{alg:T-HOSVD}.
\begin{algorithm}[!ht]\label{alg:T-HOSVD}
\caption{Truncated higher-order singular value decomposition}
\KwIn{tensor $\bm{\mathcal{T}}\in\mathbb{R}^{q_{0}\times q_{1} \times\dots\times q_{D}}$, Tucker rank $\bm{r} = (r_{0},r_{1},\dots,r_{D})$}
\Begin{
 	\For{$d = 0,1,\cdots,D$}{
		Calculate $\tilde{\bm{U}}_{d} =$ leading $r_{d}$ columns of $\operatorname{SVD}\{\mathscr{M}_{d}(\bm{\mathcal{T}})\}$\;
  	}
}
\KwOut{$\mathcal{H}_{\bm{r}}(\bm{\mathcal{T}}) = \bm{\mathcal{T}} \times_{d=0}^{D} (\tilde{\bm{U}}_{d}\tilde{\bm{U}}_{d}^{\top})$}
\end{algorithm}


\bigskip
\begin{center}
	{\large\bf SUPPLEMENTARY MATERIAL}
\end{center}

\begin{description}
	\item[supp] All proofs of the technical results and additional numerical results are collected in an online Supplementary Material. (.pdf file)
	\item[code] The code and data are made available in a GitHub repository (\url{https://github.com/kellty/FTReg}).
\end{description}

\bibliographystyle{agsm}
\bibliography{ref}
\end{document}